\newcommand{\eps}{\epsilon}

\newcommand{\sig}{{\{\sigma, \bar{\sigma}\}}}
\newcommand{\sigp}{{\{\sigma', \bar{\sigma}'\}}}

\documentclass[pre, amssymb, showkeys, showpacs, twocolumn]{revtex4}
\usepackage{graphicx}
\begin{document}

\title{Semiconservative Replication in the Quasispecies Model}

\author{Emmanuel Tannenbaum}
\email{etannenb@fas.harvard.edu}
\author{Eric J. Deeds}
\email{deeds@fas.harvard.edu}
\author{Eugene I. Shakhnovich}
\email{eugene@belok.harvard.edu}
\affiliation{Harvard University, Cambridge, MA 02138}

\begin{abstract}

This paper extends Eigen's quasispecies equations to account for the 
semiconservative nature of DNA replication.  We solve the equations in
the limit of infinite sequence length for the simplest case of a static, 
sharply peaked fitness landscape.  We show that the error catastrophe 
occurs when $ \mu $, the product of sequence length and per base pair
mismatch probability, exceeds $ 2 \ln \frac{2}{1 + 1/k} $, where $ k > 1 $ is
the first order growth rate constant of the viable ``master'' sequence (with
all other sequences having a first-order growth rate constant of $ 1 $).  This
is in contrast to the result of $ \ln k $ for conservative replication.
In particular, as $ k \rightarrow \infty $, the error catastrophe is never
reached for conservative replication, while for semiconservative replication
the critical $ \mu $ approaches $ 2 \ln 2 $.  Semiconservative replication is
therefore considerably less robust than conservative replication to the
effect of replication errors.  We also show that the mean equilibrium
fitness of a semiconservatively replicating system is given by $ k 
(2 e^{-\mu/2} - 1) $ below the error catastrophe, in contrast to the
standard result of $ k e^{-\mu} $ for conservative replication (derived
by Kimura and Maruyama in 1966).

\end{abstract}

\pacs{87.23.Kg, 87.16.Ac, 64.90.+b}

\keywords{Semiconservative, DNA, single-fitness peak, error catastrophe, 
quasispecies}

\maketitle

\section{Introduction}

In 1971, Manfred Eigen introduced the quasispecies formulation of molecular
evolution to explain the observed distribution of genotypes in RNA evolution
experiments \cite{EIG1, EIG2}.  The central result of his model was that due to
mutations, the equilibrium distribution of genotypes did not consist of 
a fittest sequence, but rather a set of closely related strains, which Eigen
termed a ``quasispecies.''  Eigen showed that a stable quasispecies only exists
if the mutation rate is kept below a threshold value.  Above this value, the
distribution of genotypes undergoes a second-order phase transition termed
the error catastrophe, in which the distribution completely delocalizes over
the gene sequence space.  Subsequent studies on the quasispecies model have
focused almost exclusively on the error catastrophe \cite{QUAS1, QUAS2,
QUAS3, QUAS4, QUAS5, QUAS6, QUAS7, QUAS8}, though there has also been some 
work on the dynamical aspects of the equations \cite{DYN1, DYN2}.  More 
recently, other phase transitions besides the error catastrophe (e.g.
the so-called ``repair catastrophe'') have been shown to arise from the 
quasispecies equations \cite{REP1, REP2}.

A common feature of previous work on the quasispecies equations has been
the implicit assumption that the genome of an organism could be written as 
a linear symbol sequence, and that replication occurs conservatively
(that is, the original genetic material is preserved during replication).
These two assumptions allow for a relatively straightforward derivation of
a system of equations modelling the evolution of a unicellular, asexual 
population.  In the simplest formulation, we assume that each organism has
a genome $ \sigma = s_1 s_2 \dots s_L $ of length $ L $, where each ``letter'' 
or ``base'' $ s_i $ is drawn from an alphabet of size $ S $ ($ = 4 $ for all 
known terrestrial life).  We assume first-order growth kinetics, and that
the genome determines the first-order growth rate constant, or fitness, denoted
by $ \kappa_{\sigma} $ (in general, $ \kappa_{\sigma} $ will be time-dependent,
reflecting the generally dynamic nature of the environment).  Furthermore,
we assume a per base replication error probability of $ \epsilon_{\sigma} $.
If we let $ x_{\sigma} $ denote the fraction of organisms with genome 
$ \sigma $, then it may be shown that \cite{EIG1, EIG2},
\begin{equation}
\frac{d x_{\sigma}}{dt} = \sum_{\sigma'}{\kappa_m(\sigma', \sigma) x_{\sigma'}}
- \bar{\kappa}(t) x_{\sigma}
\end{equation}
where $ \bar{\kappa}(t) \equiv \sum_{\sigma}{\kappa_{\sigma} x_{\sigma}} $
is simply the mean fitness of the population, and $ \kappa_m(\sigma', \sigma) $
is the first-order mutation rate constant for mutations from $ \sigma' $ to
$ \sigma $.  If $ p_m(\sigma', \sigma) $ denotes the probability of mutation
from $ \sigma' $ to $ \sigma $, then it is clear that $ \kappa_m(\sigma', 
\sigma) = \kappa_{\sigma'} p_m(\sigma', \sigma) $.  If we let $ HD(\sigma',
\sigma) $ denote the Hamming distance between $ \sigma' $ and $ \sigma $,
then it is possible to show that,
\begin{equation}
p_m(\sigma', \sigma) = (\frac{\eps_{\sigma'}}{S-1})^{HD(\sigma', \sigma)}
(1 - \eps_{\sigma'})^{L - HD(\sigma', \sigma)}
\end{equation}

The simplest formulation of these equations considers a genome-independent
replication error probability $ \eps $, and a time-independent fitness
landscape characterized by a single ``master'' sequence $ \sigma_0 $ of
fitness $ k > 1 $, with all other sequences set to a fitness of $ 1 $.  This
so-called Single Fitness Peak model (SFP) has been the subject of considerable
theoretical treatment \cite{QUAS1, QUAS2, QUAS3} (and references therein).  
The central result of this model is that, in the limit of $ L \rightarrow 
\infty $, the mean equilibrium fitness of the population is given by 
$ k e^{-\mu} $ for $ \mu \leq \ln k $, and $ 1 $ for $ \mu > \ln k $, where 
$ \mu \equiv L \eps $.  When $ \mu < \ln k $, the population is localized in a 
cluster about the master sequence, resulting in what Eigen called a 
quasispecies.  When $ \mu > \ln k $, the population is completely delocalized 
over the gene sequence space, so that no discernible quasispecies exists.  The 
transition between the two regimes, at $ \mu_{crit} \equiv \ln k $, is known 
as the error catastrophe.  It should be noted that the result of $ k e^{-\mu} $
was first derived in 1966 by Kimura and Maruyama \cite{KM66}, and is a standard
result in theoretical population genetics.

While the assumption of a linear symbol sequence and conservative replication
is correct for modelling single-stranded RNA, a proper extension of
the quasispecies model to real organisms should take into account the
double-stranded nature of DNA, and also the semiconservative nature of
DNA replication.  In semiconservative replication, the original DNA
molecule is not preserved after replication.  Rather, each strand serves
as the template for the synthesis of a complementary daughter strand, meaning
that after replication, each DNA molecule consists of one parent and one 
daughter strand \cite{VOET}. 

The formulation of the quasispecies equations given above are inadequate
to describe evolution with double-stranded, semiconservatively replicated
genomes.  There are two reasons for this:  First of all, because DNA is
double-stranded, there is no well-defined Hamming distance between two
DNA molecules.  Furthermore, because in semiconservative replication the 
original molecule is destroyed, a mathematical formulation of this process 
must incorporate an effective death term, which is clearly lacking in
the quasispecies equations for conservative replication.

The goal of this paper is to extend Eigen's formulation of the quasispecies 
equations, to account for the double-stranded and semiconservative nature of 
DNA replication.  This is a necessary first step toward making the 
quasispecies equations a quantitative tool for analyzing the evolutionary 
dynamics of unicellular organisms.  Then, after obtaining the form of Eigen's 
equations for the case of double-stranded DNA, we wish to proceed and solve 
these equations for the simplest landscape, that of the static single fitness 
peak.

This paper is organized as follows:  In the following section, we present 
an overview of DNA sequence analysis and replication mechanism, followed
by a derivation of the appropriate quasispecies equations.  We continue
in Section III with a discussion of the single fitness peak model.  
Specifically, we present the infinite sequence length equations, leaving the 
details of the derivation, which are fairly involved, for Appendix A.  We 
then go on to discuss the error catastrophe, presenting both analytical 
results and numerical corroboration using stochastic simulations of 
replicating populations.  In Section IV, we discuss our results, and also the 
extension of our equations to multiple gene models.  Finally, we conclude in 
Section V with a summary of our results, and a discussion of future research 
plans.

\section{Derivation of the Quasispecies Equations for Semiconservative 
         Replication}

\subsection{An Overview of DNA Sequence Analysis}

Double-stranded DNA consists of two anti-parallel, complementary strands.
During transcription, messenger RNA (mRNA) is synthesized in the $ 5' $ to
$ 3' $ direction.  The DNA template strand from which RNA synthesis occurs
is known as the anti-sense strand, and is read in the $ 3' $ to $ 5' $
direction.  The complementary strand, the sense strand, has the same 
sequence as the transcribed mRNA, and is ``read'' in the $ 5' $ to $ 3' $
direction (the quotes are to indicate that the sense strand does not directly
participate in the transcription process).  We therefore adopt the convention
that DNA and RNA sequences are read in the $ 5' $ to $ 3' $ direction, as
illustrated in Figure 1.  However, this convention is arbitrary, and it
is equally valid to read DNA and RNA sequences in the $ 3' $ to $ 5' $
directions.  Once a convention is adopted, the anti-parallel nature
of double-stranded DNA (or RNA) means that the complementary strands are
read in opposite directions.  A more detailed explanation can be found
in \cite{VOET}.

We consider a double-stranded DNA molecule with generalized alphabet of
size $ 2S $, consisting of ``letters'' $ 0, 1, \dots, 2S-1 $.  Each ``letter''
$ i $ is assumed to uniquely base pair with $ (i + S) \mbox{ mod } 2S $.  For 
actual DNA, we of course have $ S = 2 $, and we may make the assignment 
$ \mbox{A} \rightarrow 0 $, $ \mbox{G} \rightarrow 1 $, $ \mbox{T} 
\rightarrow 2 $, $ \mbox{C} \rightarrow 3 $.

Given a DNA molecule of sequence length $ L $, let one of the strands be
denoted by $ \sigma = b_1 \dots b_L $.  If the complement of a base $ b_i $
is denoted by $ \bar{b}_i $, then the complementary strand is given
by $ \bar{\sigma} = \bar{b}_L \dots \bar{b}_1 $.  Note that
$ \bar{\bar{\sigma}} = \sigma $, and therefore, each DNA
molecule may be denoted by the set $ \{\sigma, \bar{\sigma}\} = 
\{\bar{\sigma}, \sigma\} $.

\begin{figure}
\includegraphics[width = 0.9\linewidth]{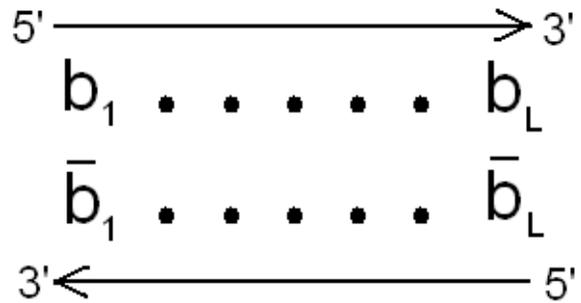}
\caption{The anti-parallel nature of double-stranded DNA and RNA}
\end{figure}

For single-stranded molecules of length $ L $ and alphabet size $ 2S $, there
are $ (2S)^L $ distinct sequences.  We seek to derive the analogous
formula for double-stranded DNA.  Given a DNA molecule $ \{\sigma, 
\bar{\sigma}\} $, define the internal Hamming distance $ l_I = HD(\sigma,
\bar{\sigma}) $.  If we let $ N_I(l_I; L) $ denote the number of DNA
molecules of length $ L $ with internal Hamming distance $ l_I $, then
the total number of distinct sequences is simply given by $ N_{tot} = 
\sum_{l_I = 0}^{L}{N_I(l_I; L)} $.  We therefore proceed to compute
$ N_I(l_I; L) $.  Due to the possibility of palindromic molecules ($ \sigma
= \bar{\sigma} $), we need to consider the case of $ L $ even and $ L $ odd
separately.

Given some DNA molecule $ \{\sigma, \bar{\sigma}\} $, with $ \sigma = 
b_1 \dots b_L $, suppose we have $ b_i = \bar{b}_{L-i+1} $ for some $ i $.  
Then $ \bar{b}_i = b_{L-i+1} $, and hence equality between corresponding bases 
in $ \sigma $ and $ \bar{\sigma} $ comes in pairs whenever $ i \neq 
L - i + 1 $.  This must always be true, since, if $ i = L - i + 1 $, then 
$ b_i = \bar{b}_{L-i+1} \Rightarrow b_i = \bar{b}_i $, which is impossible.  
Therefore, $ \sigma $ and $ \bar{\sigma} $ must be equal at an even number of 
sites, hence $ l_I $ must be odd for odd $ L $ and even for even $ L $.

Suppose $ L $ is odd, so $ L = 2l + 1 $, and consider some $ l_I = 2k + 1 $.
We have complete freedom to choose $ b_1, \dots, b_{l+1} $.  We automatically
have $ b_{l+1} \neq \bar{b}_{L - l - 1 + 1} $.  Thus, we have $ l - k $ 
remaining sites among $ b_1, \dots, b_l $ where we choose $ b_{L - i + 1} $ 
such that $ \bar{b}_{L - i + 1} = b_i $.  Equivalently, we have $ k $
sites among $ b_1, \dots, b_l $ where we choose $ b_{L - i + 1} $ such that
$ \bar{b}_{L - i + 1} \neq b_i $.  There are $ {l \choose k} $ ways of
choosing these sites, and for each such choice, there are $ 2S - 1 $ possible
values for each $ b_{L - i + 1} $ taken to be distinct from $ \bar{b}_i $.
Putting together all the degeneracies, we obtain $ {l \choose k} (2S)^{l+1}
(2S-1)^k $ ways of choosing $ \sigma $ such that $ HD(\sigma, \bar{\sigma})
= 2k + 1 = l_I $.  However, this still does not give us the set of all distinct
DNA molecules $ \{\sigma, \bar{\sigma}\} $ with internal Hamming distance
$ l_I $, for if $ \sigma \neq \bar{\sigma} $, then our counting method 
generates a given $ \{\sigma, \bar{\sigma} \} $ twice, by generating both
$ \sigma $ and $ \bar{\sigma} $.  Since $ \sigma = \bar{\sigma} $ if and
only if $ l_I = 0 $, which is impossible for odd $ L $, we have, finally,
that,
\begin{equation}
N_I(l_I; L) = \frac{1}{2}{l \choose k} (2S)^{l+1} (2S-1)^k
\end{equation}
for odd $ L $.  Thus, for odd $ L $, 
\begin{equation}
N_{tot} = \frac{1}{2} (2S)^{l+1} \sum_{k = 0}^{l}{{l \choose k} (2S-1)^k} =
\frac{1}{2} (2S)^L
\end{equation}

If $ L $ is even, then we may write $ L = 2l $.  In this case, $ l_I $ is
also even, and so $ l_I = 2k $ for some $ k = 0, \dots, l $.  We have complete
freedom to choose $ b_1, \dots, b_l $.  Proceeding as with the analysis 
above, we may show that there are $ {l \choose k} (2S)^l (2S-1)^k $ ways
of choosing $ \sigma $ so that $ HD(\sigma, \bar{\sigma}) = l_I $.  If
$ l_I \neq 0 $, we need to divide by $ 2 $ to get the set of all distinct
DNA molecules with internal Hamming distance $ l_I $.  Therefore,
\begin{equation}
N_I(l_I; L) = \left\{ 
\begin{array}{cc}
\frac{1}{2} {l \choose k} (2S)^l (2S - 1)^k & \mbox{for } k \neq 0
\\
(2S)^l & \mbox{for } k = 0
\end{array}
\right.
\end{equation}
for even $ L $.  Therefore, for even $ L $,
\begin{equation}
N_{tot} = \frac{1}{2} (2S)^L + \frac{1}{2} (2S)^{\frac{L}{2}}
\end{equation}
Note the additional $ \frac{1}{2} (2S)^{\frac{L}{2}} $ term arising from
the contribution of the palindromic sequences.

\subsection{Modelling DNA Replication}

The replication of DNA during cell division may be divided into three stages,
which are illustrated in Figure 2.

The first stage of DNA replication is strand separation, with each parent 
strand serving as a template for synthesizing the complementary daughter 
strands \cite{VOET}.  We may model this stage by writing that a given DNA 
molecule $ \{\sigma, \bar{\sigma}\} $ separates into the single-stranded 
sequences $ \sigma $ and $ \bar{\sigma} $.

As strand separation occurs, daughter strand synthesis is catalyzed 
via enzymes known as DNA replicases.  However, due to errors in the base 
pairing process, $ \sigma $ is not necessarily paired with $ \bar{\sigma} $.  
Rather, once cell division is finished, the original $ \sigma $ is paired 
with some $ \sigma' $, and similarly for $ \bar{\sigma} $.  

Each genome $ \sig $ has a characteristic replication mismatch probability 
$ \eps_\sig $ (a base-pair-independent mismatch probability is certainly a 
simplification, but it is an initial starting point).  Different genomes may 
have different replication fidelities, due to various replication error 
correction mechanisms which may or may not be functioning.  For example, in 
{\it Escherichia coli}, the DNA replicase Pol III has a built-in proofreading 
mechanism which excises mismatched bases in the daughter strand \cite{VOET}.  
In addition, in many prokaryotes and eukaryotes, DNA daughter strand synthesis 
is followed by mismatch repair \cite{VOET}, which can distinguish between the 
parent and daughter strands, thereby allowing the proper repair of 
mismatches.  All such repair mechanisms are gathered within $ \eps_\sig $ in 
our model.  

In the final stage, DNA replication and cell division is complete, and the
parent and daughter strands have become indistinguishable.  Remaining 
mismatches in the daughter cells' DNA are eliminated by various maintenance 
enzymes, which recognize and repair the lesions caused by mismatched base 
pairs.  However, because it is impossible to determine which strand has the 
incorrect base, the mismatch is correctly repaired with probability $ 1/2 $.  
The result is that the $ \sigma, \sigma' $ pair is converted to some 
$ \sigma'', \bar{\sigma}'' $, giving the DNA molecule $ \{\sigma'', 
\bar{\sigma}'' \} $.  A similar process happens for the parent 
$ \bar{\sigma} $ strand.
 
\begin{figure}
\includegraphics[width = 0.9\linewidth]{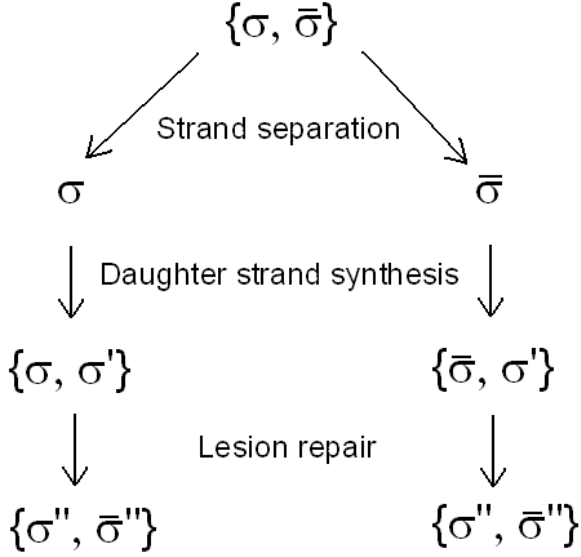}
\caption{The three stages of DNA replication}
\end{figure}

We wish to derive the probability that a given parent strand $ \sigma $ 
produces $ \{\sigma'', \bar{\sigma}''\} $ in the daughter cell.  Let us
denote this probability by $ p(\sigma, \{\sigma'', \bar{\sigma}''\}) $.  
Also, let $ p(\sigma, \sigma') $ denote the probability that $ \sigma $
is paired with $ \sigma' $ during daughter strand synthesis, and let
$ p((\sigma, \sigma'), (\sigma'', \bar{\sigma}'')) $ be the probability
that $ \sigma \rightarrow \sigma'' $, $ \sigma' \rightarrow \bar{\sigma}'' $
during post-replicative lesion repair.  Then we have, assuming $ \sigma'' 
\neq \bar{\sigma}'' $, that,
\begin{eqnarray}
p(\sigma, \{\sigma'', \bar{\sigma}''\}) 
& = & 
\sum_{\sigma'} p(\sigma, \sigma') \times \nonumber \\
&   & 
[p((\sigma, \sigma'), (\sigma'', \bar{\sigma}''))  
+ 
p((\sigma, \sigma'), (\bar{\sigma}'', \sigma''))] \nonumber \\
& = & 
\sum_{\sigma'}
p(\sigma, \sigma') p((\sigma, \sigma'), 
(\sigma'', \bar{\sigma}')) \nonumber \\
&   &
+ \sum_{\sigma'}
p(\sigma, \sigma') p((\sigma, \sigma'), (\bar{\sigma}'', \sigma''))
\end{eqnarray}
If $ \sigma'' = \bar{\sigma}'' $, then only one of the sums is kept.

We now proceed to compute $ \sum_{\sigma'}{p(\sigma, \sigma') p((\sigma,
\sigma'), (\sigma'', \bar{\sigma}''))} $.  Write $ \sigma = b_1 \dots b_L $,
$ \sigma' = b_1' \dots b_L' $, and $ \sigma" = b_1" \dots b_L" $.  Let
$ l \equiv HD(\sigma, \sigma") $.  Let us consider some $ i $ for which
$ b_i = b_i" $.  Then $ b_i' $ can take on any value, for if $ b_i' = 
\bar{b}_i" $, then no repair is necessary, and we obtain $ b_i \rightarrow 
(b_i", \bar{b}_i") $.  If $ b_i' \neq \bar{b}_i" $, then repair is necessary, 
and with probability $ 1/2 $ it is $ b_i' $ that is repaired to $ \bar{b}_i" $,
giving once again that $ b_i \rightarrow (b_i", \bar{b}_i") $.  So, let us now 
consider some $ i $ for which $ b_i \neq b_i" $.  Then $ b_i' $ must be equal 
to $ \bar{b}_i" $.  Otherwise, if $ b_i' = \bar{b}_i \neq \bar{b}_i" $, then 
no lesion repair occurs, and we get $ b_i \rightarrow (b_i, \bar{b}_i) \neq 
(b_i", \bar{b}_i") $.  If $ b_i' \neq \bar{b}_i $, then $ b_i' $ is repaired 
with probability $ 1/2 $ to $ \bar{b}_i $, or $ b_i $ is repaired with 
probability $ 1/2 $ to $ \bar{b}_i' $.  Thus, either $ b_i \rightarrow (b_i, 
\bar{b}_i) \neq (b_i", \bar{b}_i") $, or $ b_i \rightarrow (\bar{b}_i', b_i') 
\neq (b_i", \bar{b}_i") $, and so the corresponding $ \sigma' $ does not 
contribute to the sum $ \sum_{\sigma'}{p(\sigma, \sigma') p((\sigma, \sigma'), 
(\sigma", \bar{\sigma}"))} $, since $ p((\sigma, \sigma'), (\sigma", 
\bar{\sigma}")) = 0 $.  

Our analysis allows us to perform the sum, assuming a probability 
$ \eps_\sig $ of a mismatch.  For a given $ \sigma' $, let $ l' $ denote 
the number of sites among the $ L-l $ sites which are equal in $ \sigma $ 
and $ \sigma" $ for which $ b_i' \neq \bar{b}_i" $.  There are 
$ {{L-l} \choose l'} (2S-1)^{l'} $ ways of choosing such a $ \sigma' $.  The 
probability $ p(\sigma, \sigma') $ is equal to $ (\frac{\eps_\sig}{2S-1})^
{l'+l} (1 - \eps_\sig)^{L-l-l'} $.  The probability $ p((\sigma, \sigma'), 
(\sigma", \bar{\sigma}")) $ is then $ (\frac{1}{2})^{l'+l} $, so multiplying 
by the degeneracy in $ l' $ and summing over all $ l' $ gives, 
\begin{widetext}
\begin{eqnarray}
\sum_{\sigma'}{p(\sigma, \sigma')p((\sigma, \sigma'), 
(\sigma", \bar{\sigma}"))} & = &
\sum_{l' = 0}^{L-l}{{{L-l} \choose l'} (2S-1)^{l'} (\frac{\eps_\sig}{2S-1})^
{l'+l} (1 - \eps_\sig)^{L-l-l'} (\frac{1}{2})^{l'+l}} \nonumber \\
& = &
(\frac{\eps_\sig/2}{2S-1})^{l} (1 - \frac{\eps_\sig}{2})^{L-l}
\end{eqnarray}
If we define $ \bar{l} = HD(\sigma, \bar{\sigma}") $, then we obtain that
$ \sum_{\sigma'}{p(\sigma, \sigma')p((\sigma, \sigma'), (\bar{\sigma}", 
\sigma"))} = (\frac{\eps_\sig/2}{2S-1})^{\bar{l}} (1 - \frac{\eps_\sig}{2})^
{L - \bar{l}} $.  Now, note that $ HD(b_1 \dots b_L, b_1' \dots b_L') = 
HD(\bar{b}_L \dots \bar{b}_1, \bar{b}_L' \dots \bar{b}_1') $, and 
$ HD(b_1 \dots b_L, \bar{b}_L' \dots \bar{b}_1') = HD(\bar{b}_L \dots 
\bar{b}_1, b_1' \dots b_L') $, so that $ l = HD(\sigma, \sigma") =
HD(\bar{\sigma}, \bar{\sigma}") $, and $ \bar{l} = HD(\sigma, \bar{\sigma}") = 
HD(\bar{\sigma}, \sigma") $.  Therefore, we obtain that,
\begin{equation}
p(\sigma, \{\sigma", \bar{\sigma}"\}) = p(\bar{\sigma}, 
\{\sigma", \bar{\sigma}"\}) = \left\{\begin{array}{cc}
(\frac{\eps_\sig/2}{2S-1})^{l}(1 - \frac{\eps_\sig}{2})^{L-l} +
(\frac{\eps_\sig/2}{2S-1})^{\bar{l}}(1 - \frac{\eps_\sig}{2})^{L - \bar{l}} & 
\mbox{for } \sigma" \neq \bar{\sigma}" \\
(\frac{\eps_\sig/2}{2S-1})^{l}(1 - \frac{\eps_\sig}{2})^{L-l} & 
\mbox{for } \sigma" = \bar{\sigma}"
\end{array}
\right.
\end{equation}
\end{widetext}

\subsection{The Quasispecies Equations}

We are now ready to derive the quasispecies equations for semiconservative
replication.  We consider a population of unicellular, asexually replicating
organisms.  Let $ n_{\{\sigma, \bar{\sigma}\}} $ denote the number of 
organisms with genome $ \{\sigma, \bar{\sigma}\} $.  We let $ \kappa_{\{\sigma,
\bar{\sigma}\}} $ denote the first-order growth rate constant of organisms
with genome $ \{\sigma, \bar{\sigma}\} $.  Then from the replication mechanism
illustrated in Figure 2, we obtain the system of differential equations
given by,
\begin{eqnarray}
\frac{d n_{\{\sigma, \bar{\sigma}\}}}{dt} 
& = &
-\kappa_{\{\sigma, \bar{\sigma}\}} n_{\{\sigma, \bar{\sigma}\}} \nonumber \\
&   & 
+ \sum_{\sigp}{\kappa_{\sigp} n_{\sigp}} \times \nonumber \\
&   &
(p(\sigma', \sig) + p(\bar{\sigma}', \sig))
\end{eqnarray}
The first term is a death term which takes into account the destruction
of the original genome during replication.  The terms in the summation
take into the account the production of $ \sig $ from both $ \sigma' $ and
$ \bar{\sigma}' $.

We now define $ n = \sum_{\sig}{n_{\sig}} $, and $ x_{\sig} = n_{\sig}/n $.
Reexpressed in terms of the population fractions $ x_{\sig} $, the
dynamical equations become,
\begin{eqnarray}
\frac{d x_{\sig}}{dt} 
& = & 
\sum_{\sigp} \kappa_{\sigp} x_{\sigp} \times \nonumber \\
&   &(p(\sigma', \sig) + p(\bar{\sigma}', \sig)) \nonumber \\
&   & - (\kappa_{\sig} + \bar{\kappa}(t)) x_{\sig}
\end{eqnarray}
where $ \bar{\kappa}(t) \equiv \sum_{\sig}{\kappa_{\sig} x_{\sig}} $, so is
simply the mean fitness of the population, which arises as a normalization
term to ensure that the total population fraction remains $ 1 $.

We now proceed to put these equations into a form which is more easily
amenable to analysis than the above equations.  To this end, we make
the following definitions:  (1) $ \kappa_{\sigma} \equiv \kappa_{\sig} $, so 
that $ \kappa_{\bar{\sigma}} = \kappa_{\sigma} $.  (2) $ \eps_{\sigma} \equiv 
\eps_\sig $, so that $ \eps_{\sigma} = \eps_{\bar{\sigma}} $.  Finally,
$ y_{\sigma} \equiv \frac{1}{2} x_\sig $ if $ \sigma \neq \bar{\sigma} $,
and $ y_{\sigma} \equiv x_{\sig} $ if $ \sigma = \bar{\sigma} $.  Clearly,
we also have that $ y_{\sigma} = y_{\bar{\sigma}} $.

Now,
\begin{widetext}
\begin{eqnarray}
\sum_{\sigp}{\kappa_{\sigp} x_{\sigp} (p(\sigma', \sig) + p(\bar{\sigma}',
\sig))} 
& = & 
\sum_{\sigp, \sigma' \neq \bar{\sigma}'}{\kappa_{\sigp} x_{\sigp}}
(p(\sigma', \sig) + p(\bar{\sigma}', \sig)) \nonumber \\
&   &
+ \sum_{\sigp, \sigma' = \bar{\sigma}'}{\kappa_{\sigp} x_{\sigp}}
(p(\sigma', \sig) + p(\bar{\sigma}', \sig)) \nonumber \\
& = & 
\sum_{\sigp, \sigma' \neq \bar{\sigma}'}{\kappa_{\sigma'} x_{\sigp}
p(\sigma', \sig)} \nonumber \\
&   &
+ \sum_{\sigp, \sigma' \neq \bar{\sigma}'}
{\kappa_{\bar{\sigma}'} x_{\sigp} p(\bar{\sigma}', \sig)} \nonumber \\
&   &
+ \sum_{\sigp, \sigma' = \bar{\sigma}'}{2 \kappa_{\sigma'} x_{\sigp}
p(\sigma', \sig)} \nonumber \\
& = & 
\sum_{\sigma', \sigma' \neq \bar{\sigma}'}{\kappa_{\sigma'}
x_{\sigp} p(\sigma', \sig)} + 2\sum_{\sigma', \sigma' = \bar{\sigma}'}
{\kappa_{\sigma'} x_{\sigp} p(\sigma', \sig)} \nonumber \\
\end{eqnarray}

Therefore, for $ \sigma \neq \bar{\sigma} $,
\begin{eqnarray}
\frac{d y_{\sigma}}{dt} = \frac{1}{2} \frac{d x_{\sig}}{dt} 
& = &
\frac{1}{2} \sum_{\sigma', \sigma' \neq \bar{\sigma}'}
{2 \kappa_{\sigma'} y_{\sigma'}} ((\frac{\eps_{\sigma'}/2}{2S-1})^
{HD(\sigma, \sigma')} 
(1 - \frac{\eps_{\sigma'}}{2})^{L - HD(\sigma, \sigma')} + 
(\frac{\eps_{\sigma'}/2}{2S-1})^{HD(\bar{\sigma}, \sigma')}
(1 - \frac{\eps_{\sigma'}}{2})^{L - HD(\bar{\sigma}, \sigma')}) \nonumber \\
&   &
+ \sum_{\sigma', \sigma' = \bar{\sigma}'}{\kappa_{\sigma'} y_{\sigma'}
((\frac{\eps_{\sigma'}/2}{2S-1})^{HD(\sigma, \sigma')} 
(1 - \frac{\eps_{\sigma'}}{2})^{L - HD(\sigma, \sigma')} + 
(\frac{\eps_{\sigma'}/2}{2S-1})^{HD(\bar{\sigma}, \sigma')} 
(1 - \frac{\eps_{\sigma'}}{2})^{L - HD(\bar{\sigma}, \sigma')})} \nonumber \\
&   &
- (\kappa_{\sigma} + \bar{\kappa}(t)) y_{\sigma} \nonumber \\
& = &
\sum_{\sigma'}{\kappa_{\sigma'} y_{\sigma'}}
((\frac{\eps_{\sigma'}/2}{2S-1})^{HD(\sigma, \sigma')} 
(1 - \frac{\eps_{\sigma'}}{2})^{L - HD(\sigma, \sigma')} + 
(\frac{\eps_{\sigma'}/2}{2S-1})^{HD(\bar{\sigma}, \sigma')} 
(1 - \frac{\eps_{\sigma'}}{2})^{L - HD(\bar{\sigma}, \sigma')}) \nonumber \\
&   &
- (\kappa_{\sigma} + \bar{\kappa}(t)) y_{\sigma} \nonumber \\
& = & 
\sum_{\sigma'}{\kappa_{\sigma'} y_{\sigma'} 
(\frac{\eps_{\sigma'}/2}{2S-1})^{HD(\sigma, \sigma')} 
(1 - \frac{\eps_{\sigma'}}{2})^{L - HD(\sigma, \sigma')}} + 
\sum_{\sigma'}{\kappa_{\bar{\sigma}'} y_{\bar{\sigma}'} 
(\frac{\eps_{\bar{\sigma}'}/2}{2S-1})^{HD(\sigma, \bar{\sigma}')} 
(1 - \frac{\eps_{\bar{\sigma}'}}{2})^{L - HD(\sigma, \bar{\sigma}')}} 
\nonumber \\
&   & - (\kappa_{\sigma} + \bar{\kappa}(t)) y_{\sigma} \nonumber \\
& = & 2 \sum_{\sigma'}{\kappa_{\sigma'} y_{\sigma'} 
(\frac{\eps_{\sigma'}/2}{2S-1})^{HD(\sigma, \sigma')}
(1 - \frac{\eps_{\sigma'}}{2})^{L - HD(\sigma, \sigma')}} - 
(\kappa_{\sigma} + \bar{\kappa}(t)) y_{\sigma}  
\end{eqnarray}

For $ \sigma = \bar{\sigma} $ we get,
\begin{eqnarray}
\frac{d y_{\sigma}}{dt} = \frac{d x_{\sig}}{dt} & = &
\sum_{\sigma', \sigma' \neq \bar{\sigma}'}{2 \kappa_{\sigma'} y_{\sigma'}
p(\sigma', \sig)} + \nonumber \\ 
&   & 
2 \sum_{\sigma', \sigma' = \bar{\sigma}'}{\kappa_{\sigma'} 
y_{\sigma'} p(\sigma', \sig)} - (\kappa_{\sigma} + \bar{\kappa}(t)) y_{\sigma}
\nonumber \\
& = & 
2 \sum_{\sigma'}{\kappa_{\sigma'} y_{\sigma'} (\frac{\eps_{\sigma'}/2}{2S-1})^
{HD(\sigma, \sigma')}
(1 - \frac{\eps_{\sigma'}}{2})^{L - HD(\sigma, \sigma')} - (\kappa_{\sigma} + 
\bar{\kappa}(t)) y_{\sigma}}
\end{eqnarray}
\end{widetext}

Since we obtain the same set of equations for palindromic and non-palindromic 
molecules, the final form of our quasispecies equations becomes,
\begin{eqnarray}
\frac{d y_{\sigma}}{dt} 
& = & 
2 \sum_{\sigma'}{\kappa_{\sigma'} y_{\sigma'}
(\frac{\eps_{\sigma'}/2}{2S-1})^{HD(\sigma, \sigma')} 
(1 - \frac{\eps_{\sigma'}}{2})^{L - HD(\sigma, \sigma')}} \nonumber \\
&   & 
- (\kappa_{\sigma} + \bar{\kappa}(t)) y_{\sigma}
\end{eqnarray}
It is readily shown that $ \bar{\kappa}(t) = \sum_{\sigma}{\kappa_{\sigma} 
y_{\sigma}} $.  It is also readily shown that $ y_{\sigma} = y_{\bar{\sigma}} 
$ for all $ \sigma $ implies that $ d y_{\sigma}/dt = d y_{\bar{\sigma}}/dt $ 
for all $ \sigma $, and so $ y_{\sigma} = y_{\bar{\sigma}} $ is preserved by 
the evolution.

\section{The Single Fitness Peak}

\subsection{Overview and Analytical Results}

In the single fitness peak model, there exists a unique, master genome
$ \{\sigma_0, \bar{\sigma}_0\} $ with fitness $ k > 1 $, with all other 
genomes having fitness $ 1 $.  Our fitness landscape is therefore given
by $ \kappa_{\sigma_0} = \kappa_{\bar{\sigma}_0} = k $, while $ \kappa_{\sigma}
= 1 $ for $ \sigma \neq \sigma_0, \bar{\sigma}_0 $.  We also assume
that $ \eps_{\sigma} $ is independent of $ \sigma $, so that $ \eps_{\sigma}
= \eps $.  For this landscape, we wish to obtain the equilibrium behavior of 
the system of differential equations given by Eq. (15).

For the case of conservative replication, the single fitness peak model may
be solved by first grouping the genomes into Hamming classes \cite{QUAS1}.  
Specifically, given the master sequence $ \sigma_0 $, we may define $ HC(l) =
\{\sigma| HD(\sigma, \sigma_0) = l\} $.  If $ x_{\sigma} $ denotes the
population fraction with genome $ \sigma $, then we define $ z_l =
\sum_{\sigma \in HC(l)}{x_{\sigma}} $.  The quasispecies equations are then 
reexpressed in terms of the $ z_l $, and the equilibrium equations may be 
readily solved in the limit of infinite sequence length, since the 
backmutation terms become negligible.  The result is,
\begin{equation}
\frac{d z_l}{dt} = e^{-\mu} \sum_{l_1 = 0}^{l}{\frac{1}{l_1!} \kappa_{l-l_1}
\mu^{l_1} z_{l-l_1}} - \bar{\kappa}(t) z_l
\end{equation}
where $ \kappa_l = k $ for $ l = 0 $, and $ 1 $ for $ l > 0 $, 
$ \bar{\kappa}(t) = (k-1) z_0 + 1 $, and $ \mu \equiv L \eps $ in the limit 
$ L \rightarrow \infty $.

For the case of semiconservative replication, the single fitness peak model
for double-stranded genomes becomes an effectively two fitness peak model.  
Thus, it is not possible to directly group the genomes into Hamming classes.  
Nevertheless, the single fitness peak for double-stranded genomes is 
solvable.  The details of the solution, which are fairly involved, may
be found in Appendix A.  The final result, however, is simple to understand.
In the limit of infinite sequence length, $ \sigma_0 $ and $ \bar{\sigma}_0 $
become infinitely separated.  Therefore, locally around $ \sigma_0 $,
$ \bar{\sigma}_0 $ we have an effectively single fitness peak model.  We
may therefore exploit the local symmetry of the landscape and define 
Hamming classes around $ \sigma_0 $ and $ \bar{\sigma}_0 $.  Thus,
$ HC(\sigma_0; l) \equiv \{\sigma| HD(\sigma, \sigma_0) = l\} $, and similarly
for $ \bar{\sigma}_0 $.  We may then define $ w_l = 
\sum_{\sigma \in HC(\sigma_0; l)}{y_{\sigma}} $, and $ \bar{w}_l $ may
be defined similarly with respect to $ \bar{\sigma}_0 $.  However, by
symmetry of the landscape we have $ w_l = \bar{w}_l $, and so need only
consider the dynamics of the $ w_l $.  In Appendix A, we show that when
expressed in terms of the $ w_l $, the quasispecies equations become,
\begin{equation}
\frac{d w_l}{dt} = 2 e^{-\mu/2} \sum_{l_1 = 0}^{l}{\frac{1}{l_1!}
(\frac{\mu}{2})^{l_1} \kappa_{l - l_1} w_{l - l_1}} - (\kappa_l +
\bar{\kappa}(t)) w_l
\end{equation}
In this case, $ \bar{\kappa}(t) = 2(k-1) w_0 + 1 $.  The reason for this
is that $ w_0 $ is only the fraction of the population on the fitness peak
at $ \sigma_0 $.  By the way we defined our $ y_{\sigma} $, the total
fraction of viable organisms is given by $ w_0 + \bar{w}_0 = 2 w_0 $.

We begin the solution of the infinite sequence length equations by solving
for the equilibrium value of $ w_0 $.  We have,
\begin{equation}
\frac{d w_0}{dt} = 2 k e^{-\mu/2} w_0 - (k + \bar{\kappa}(t)) w_0
\end{equation}
which admits the solutions $ w_0 = 0, \frac{k (2 e^{-\mu/2} - 1) - 1}{2(k-1)} 
$.  Multiplying by $ 2 $, we get the equilibrium solution for $ x_{\{\sigma_0,
\bar{\sigma}_0\}} $ of $ 0 $ or $ \frac{k (2 e^{-\mu/2} - 1) - 1}{k - 1} $.
To determine the domain of validity of these solutions, we note that
we want $ w_0 = 1/2 $ for $ \mu = 0 $.  That is, when replication is perfect,
then the population resides entirely on the fitness peak $ \{\sigma_0,
\bar{\sigma}_0 \} $.  We must also have $ w_0 \geq 0 $, which holds as
long as $ k(2 e^{-\mu/2} - 1) - 1 \geq 0 \Rightarrow
\mu \leq \mu_{crit} \equiv 2 \ln \frac{2}{1 + 1/k} $.  Therefore,
by continuity, we have that for $ \mu \leq \mu_{crit} $, the equilibrium
solution is $ w_0 = \frac{k (2 e^{-\mu/2} - 1) - 1}{2(k-1)} $.  For
$ \mu > \mu_{crit} $, the equilibrium solution becomes $ w_0 = 0 $.  
The transition between these two solutions regimes is known as the
error catastrophe.   

In dealing with conservative replication, another parameter of interest
which we consider is the localization length, defined as $ \langle l \rangle
= \sum_{l = 1}^{\infty}{l z_l} $, where $ z_l $ denotes the population
fraction at Hamming distance $ l $ from the master sequence.  We wish to 
extend the definition of localization length to our model.  The complication
here is that in the limit of infinite sequence length, the Hamming distances
$ l $ and $ \bar{l} $ to $ \sigma_0 $ and $ \bar{\sigma}_0 $ (respectively) 
cannot be simultaneously finite.  However, as mentioned previously, the 
fraction of the population at a Hamming distance $ l $ from $ \sigma_0 $, 
given by $ w_l $, is equal to the fraction of the population at a Hamming 
distance $ l $ from $ \bar{\sigma}_0 $, given by $ \bar{w}_l $.  Therefore, 
an appropriate definition for the Hamming distance is to define $ \langle l 
\rangle = \sum_{l = 1}^{\infty}{2 l w_l} $.  We may compute $ \langle l 
\rangle $ by using a technique similar to the one developed in \cite{REP2}.  
Briefly, a differential equation for the time evolution of $ \langle l 
\rangle $ is derived from the evolution equations for the $ w_l $.  The 
result is,
\begin{equation}
\frac{d \langle l \rangle}{dt} = \langle l \rangle (1 - \bar{\kappa}(t)) +
\mu \bar{\kappa}(t)
\end{equation}
giving at equilibrium that,
\begin{equation}
\langle l \rangle = \mu \frac{\bar{\kappa}(t = \infty)}
{\bar{\kappa}(t = \infty) - 1} = \mu \frac{k (2 e^{-\mu/2} - 1)}
{k (2 e^{-\mu/2} - 1) - 1}
\end{equation}
Note that the localization length is finite for $ \mu < \mu_{crit} $, but
diverges at the error catastrophe.

For convenience, Table I illustrates the difference between conservative
and semiconservative replication.

\begin{table}
\begin{tabular}{ccc}
\underline{Parameter} & \underline{Conservative} & 
\underline{Semiconservative} \\
$ \mu_{crit} $ & $ \ln k $ & $ 2 \ln \frac{2}{1 + 1/k} $ \\
$ x_{master} $ ($ \mu < \mu_{crit} $) & $ \frac{k e^{-\mu} - 1}{k-1} $ &
$ \frac{k (2 e^{-\mu/2} - 1) - 1}{k-1} $ \\
$ \bar{\kappa}(t = \infty) $ ($ \mu < \mu_{crit} $) & 
$ k e^{-\mu} $ & $ k (2 e^{-\mu/2} - 1) $ \\
$ \langle l \rangle $ ($ \mu < \mu_{crit} $) & $ \mu \frac{k e^{-\mu}}
{k e^{-\mu} - 1} $ & $ \mu \frac{k (2 e^{-\mu/2} - 1)}
{k (2 e^{-\mu/2} - 1) - 1} $ \\
\end{tabular}
\caption{Comparison of quasispecies equilibrium between conservative and
semiconservative replication.  It should be noted that 
$ \bar{\kappa}(t = \infty) $ is simply the equilibrium mean fitness of
the population.}

\end{table}

Figure 3 shows a plot of $ \mu_{crit} $ versus $ k $ for both the conservative
and semiconservative cases.  Figure 4 shows a plot of 
$ \bar{\kappa}(t = \infty) $ versus $ \mu $ for $ k = 10 $ for both the 
conservative and semiconservative cases.  Finally, Figure 5 shows a plot 
of $ \langle l \rangle $ versus $ \mu $ for $ k = 10 $ for both the 
conservative and semiconservative cases.

\begin{figure}
\includegraphics[width = 0.9\linewidth, angle = -90]{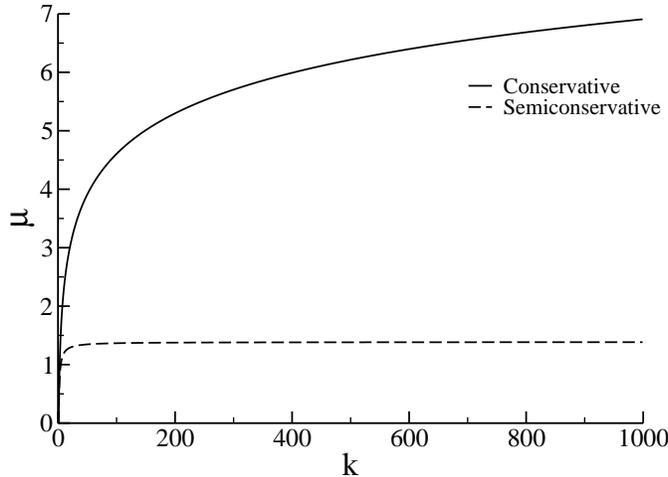}
\caption{Plot of $ \mu_{crit} $ versus $ k $ for both conservative and
semiconservative replication.}
\end{figure}

\begin{figure}
\includegraphics[width = 0.9\linewidth, angle = -90]{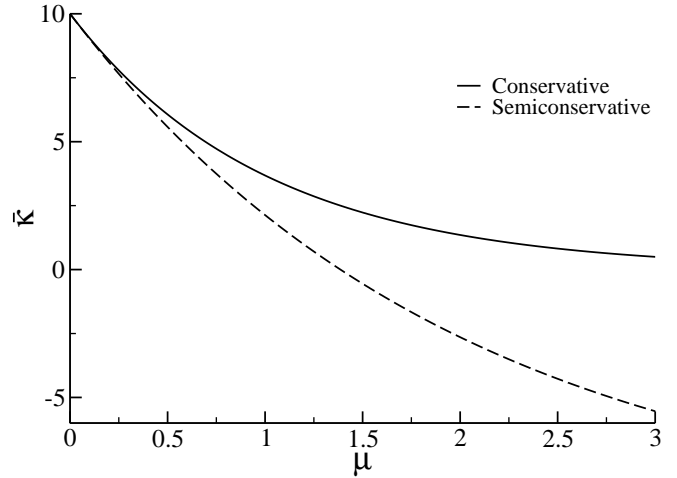}
\caption{Plot of $ \bar{\kappa}(t = \infty) $ versus $ \mu $ for
$ k = 10 $.}
\end{figure}

\begin{figure}
\includegraphics[width = 0.9\linewidth, angle = -90]{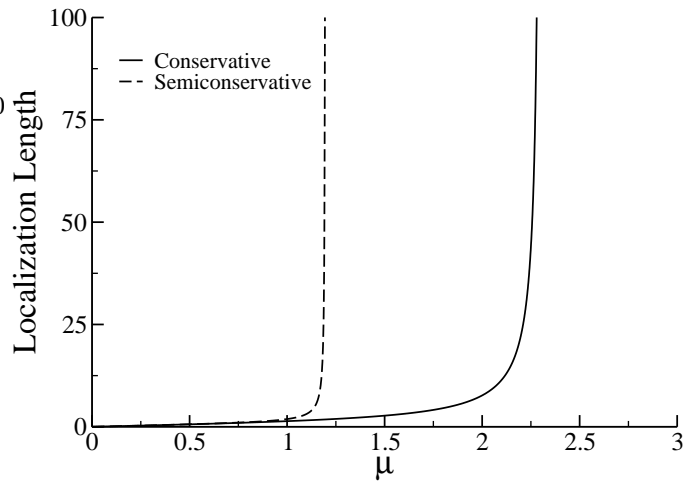}
\caption{Plot of $ \langle l \rangle $ versus $ \mu $ for
$ k = 10 $.}
\end{figure}

\subsection{Numerical Verification Using Stochastic Simulations}

In order to allow for independent verification of the semiconservative error 
catastrophe, we employ stochastic simulations of evolving organisms rather 
than numerical integration of the equations themselves.  The details of these 
simulations are given in Appendix B.

Our simulations involve constant population sizes of $ 10^4 $ organisms with 
genome lengths of $ 101 $ base pairs, using an alphabet size $ 2S = 4 $ to 
correspond with the alphabet size of DNA.  The master sequence on our SFP 
landscape replicates at each time step with probability 
$ p_{R,\{\sigma_0, \bar{\sigma}_0\}} = 1 $; all other sequences replicate with 
$ p_{R, \{\sigma, \bar{\sigma}\} \neq \{\sigma_0, \bar{\sigma}_0\}} = 0.01 $.
Transforming these replication probabilities to replication rates for use in 
the above equations, we have $ \kappa_{\{\sigma_0, \bar{\sigma}_0\}} = 100 $ 
given $ \kappa_{\{\sigma, \bar{\sigma}\} \neq \{\sigma_0, 
\bar{\sigma}_0\}} = 1 $.

We run our simulations using the above parameters.  For those simulations in 
which the equilibrium value of $ x_{\mbox{master}} > 0 $, we calculate the 
equilibrium fraction of viable organisms as the average of $ 100 $ equilibrium 
steps from $ 10 $ independent runs.  For those simulations in which the 
equilibrium value of $ x_{\mbox{master}} = 0 $, we verify this behavior in 
two independent runs.  The equilibrium value of $ x_{\mbox{master}} $ is 
calculated as above for various values of $ \epsilon $, and the results are 
displayed in Figure 6.  The predicted $ \epsilon_{crit} $ for the above 
parameters is indicated in Figure 6 as a dashed line.  Note the good agreement 
between the theoretical prediction of the error catastrophe and the 
numerical results.

\begin{figure}
\includegraphics[width = 0.9\linewidth, angle = -90]{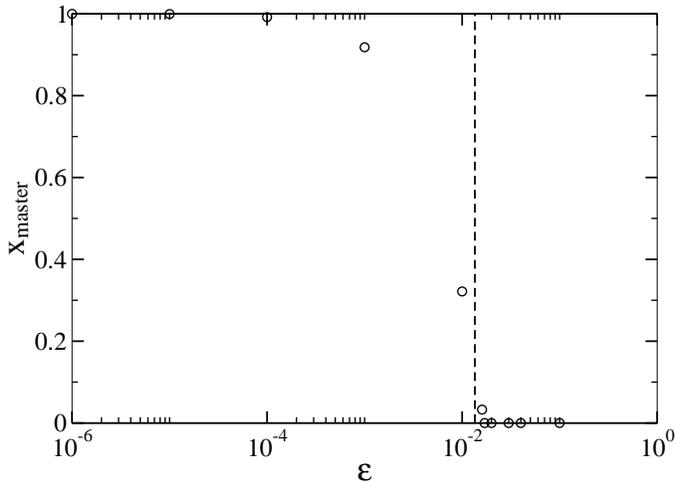}
\caption{Error catastrophe in a stochastic simulation of a finite
population of semiconservatively replicating organisms.}
\end{figure}

\section{Discussion}

The key difference between conservative replication and semiconservative 
replication is the destruction of the parent genome in the semiconservative
case, as opposed to its preservation in the conservative case.  This is 
captured by the functions $ e^{-\mu} $ versus $ 2 e^{-\mu/2} - 1 $ in the 
formulas given in Table 1.  For conservative replication, $ e^{-\mu} $ is 
simply the probability of correct replication.  This probability is always 
positive, and so, by making $ k $ sufficiently large, it is possible to 
guarantee that the effective growth rate $ k e^{-\mu} $ of the master sequence 
stays above the growth rate of $ 1 $ for the unviable sequences.  For 
semiconservative replication, the probability that each strand is matched 
with its proper complementary strand is $ e^{-\mu/2} $.  Therefore, since 
there are two parent strands, and the parent genome is destroyed during 
replication, we have the factor $ 2 e^{-\mu/2} - 1 $, yielding an effective 
growth rate of $ k (2 e^{-\mu/2} - 1) $.  However, $ 2 e^{-\mu/2} - 1 $ is 
only positive when $ e^{-\mu/2} > 1/2 $, or when $ \mu < 2 \ln 2 $.  When the 
probability of correct daughter strand synthesis drops below $ 1/2 $, then 
the rate of production of viable genomes no longer exceeds the rate of 
destruction.  The result is that replicating faster simply increases the rate 
of destruction of viable organisms, and therefore does not avoid the error 
catastrophe.

The semiconservative quasispecies formalism may be naturally extended to
more sophisticated models with more than one gene.  In this paper, we focused
on the single fitness peak model, in which the genome consists of a single,
``viability gene,'' and the replication error probability is 
genome-independent.

As an example, we may incorporate mismatch repair into the semiconservative,
quasispecies formalism.  As with the conservative case \cite{REP1, REP2}, we
consider a two gene model, in which one gene codes for viability, and the
other codes for repair.  Thus, a given genome $ \{\sigma, \bar{\sigma}\} $ 
may be written as $ \{\sigma_{via} \sigma_{rep}, \bar{\sigma}_{rep} 
\bar{\sigma}_{via} \} $.  As was done in \cite{REP1, REP2}, we may assume 
a single-fitness peak in both the viability and repair genes, so that there 
exist ``master'' sequences $ \sigma_{via, 0}, \bar{\sigma}_{via, 0} $, and 
$ \sigma_{rep, 0}, \bar{\sigma}_{rep, 0} $ for both viability and 
repair, respectively.  In the single-stranded formulation of the 
semiconservative model, a given $ \sigma $ has a first-order growth rate 
$ k > 1 $ if $ \sigma = \sigma_{via, 0} \sigma_{rep} $ or $ \bar{\sigma}_{rep} 
\bar{\sigma}_{via, 0} $.  The growth rate constant is $ 1 $ otherwise.
Furthermore, $ \sigma $ has a functioning mismatch repair system with 
failure probability $ \eps_r $ if $ \sigma = \sigma_{via} \sigma_{rep, 0} $,
or $ \bar{\sigma}_{rep, 0} \bar{\sigma}_{via} $.  Otherwise, mismatch repair
is inactivated. 

While we leave the solution of this two gene model for future work, we may
nevertheless compute the location of the repair catastrophe.  As with the
case for conservative replication, the repair catastrophe occurs when
the effective growth rate constant of viable repairers drops below the growth 
rate of constant of viable non-repairers.  For viable repairers, the effective
growth rate constant is $ k (2 e^{-\eps_r \mu/2} - 1) $.  We have for the 
non-repairers an effective growth rate constant of viable organisms given 
by $ k (2 e^{-(L_{via}/L) \mu/2} - 1) $.  The factor of $ L_{via}/L $ arises 
because in dealing with the overall growth rate of the mutators, we are only 
concerned with the production of viable organisms.  The repairer gene does 
not need to be correctly replicated.

The repair catastrophe then occurs when $ k(2 e^{-\eps_r \mu/2} - 1) = 
k (2 e^{-(L_{via}/L) \mu/2} - 1) $, or when $ \eps_r = L_{via}/L $.  
Interestingly, this result is unchanged from the point-mutation, conservative 
result in \cite{REP1}, or the full solution, conservative result in 
\cite{REP2}.

\section{Conclusions}

This paper extended the quasispecies formalism to include the case of
semiconservative replication, in order to allow for the more realistic 
modelling of the evolutionary dynamics of DNA-based life.  While our model 
is currently most directly applicable to prokaryotic genomes, which generally 
consist of a single, circular DNA molecule, we believe that it forms the 
basis for future extension to genomes consisting of multiple chromosomes.

After deriving the quasispecies equations for semiconservative systems,
we proceeded to solve them for the simplest landscape, that of the static
single fitness peak.  As with conservative replication, the solution of
the single fitness peak yielded two regimes:  A viable regime, where the
population is localized about the ``master'' genome, and an unviable regime,
where the population is delocalized over the genome space.  The transition
between the two regimes is known as the error catastrophe.

The main difference between conservative and semiconservative replication is
that for conservative replication, it is possible to push the error catastrophe
to arbitrarily high replication error rates by increasing the growth rate
constant of the master genome.  In semiconservative replication, on the
other hand, the probability of correct replication must always be greater
than $ 1/2 $, in order to avoid the error catastrophe.

Semiconservative replication is therefore considerably less robust to
the effect of mutagens than conservative replication.  Furthermore, 
the existence of a lower bound to semiconservative replication fidelity
explains why above the error catastrophe, mutagenic agents kill more rapidly
replicating cells faster than more slowly replicating cells.  Thus, our model
provides a mathematical basis for explaining the efficacy of chemotherapeutic
agents in treating cancers.

\begin{acknowledgments}

This research was supported by an NIH postdoctoral research
fellowship, and by a Howard-Hughes Medical Institute pre-doctoral research
fellowship.

\end{acknowledgments}

\begin{appendix}

\section{Solution of the Static Single Fitness Peak Model for Semiconservative
Replication}

\subsection{Finite Genome Size Equations}

To begin, let us define the internal Hamming distance $ l_I = HD(\sigma_0,
\bar{\sigma}_0) $.  Also, let $ \sigma_{0,S} $ denote the subsequence of
bases where $ \sigma_0 $ and $ \bar{\sigma}_0 $ are identical, and 
$ \sigma_{0, NS} $ and $ \bar{\sigma}_{0, NS} $ denote the subsequences of 
bases in $ \sigma_0 $ and $ \bar{\sigma}_0 $, respectively, where they differ.
Then given some gene sequence $ \sigma $, we can break it up into two
subsequences $ \sigma_S $ and $ \sigma_{NS} $.  $ \sigma_S $ denotes the
subsequence of bases in $ \sigma $ corresponding to the subsequence of bases 
where $ \sigma_0 $, $ \bar{\sigma}_0 $ are identical.  $ \sigma_{NS} $ denotes
the subsequence of remaining bases.  This is illustrated in Figure 3.

\begin{figure}
\includegraphics[width = 0.9\linewidth]{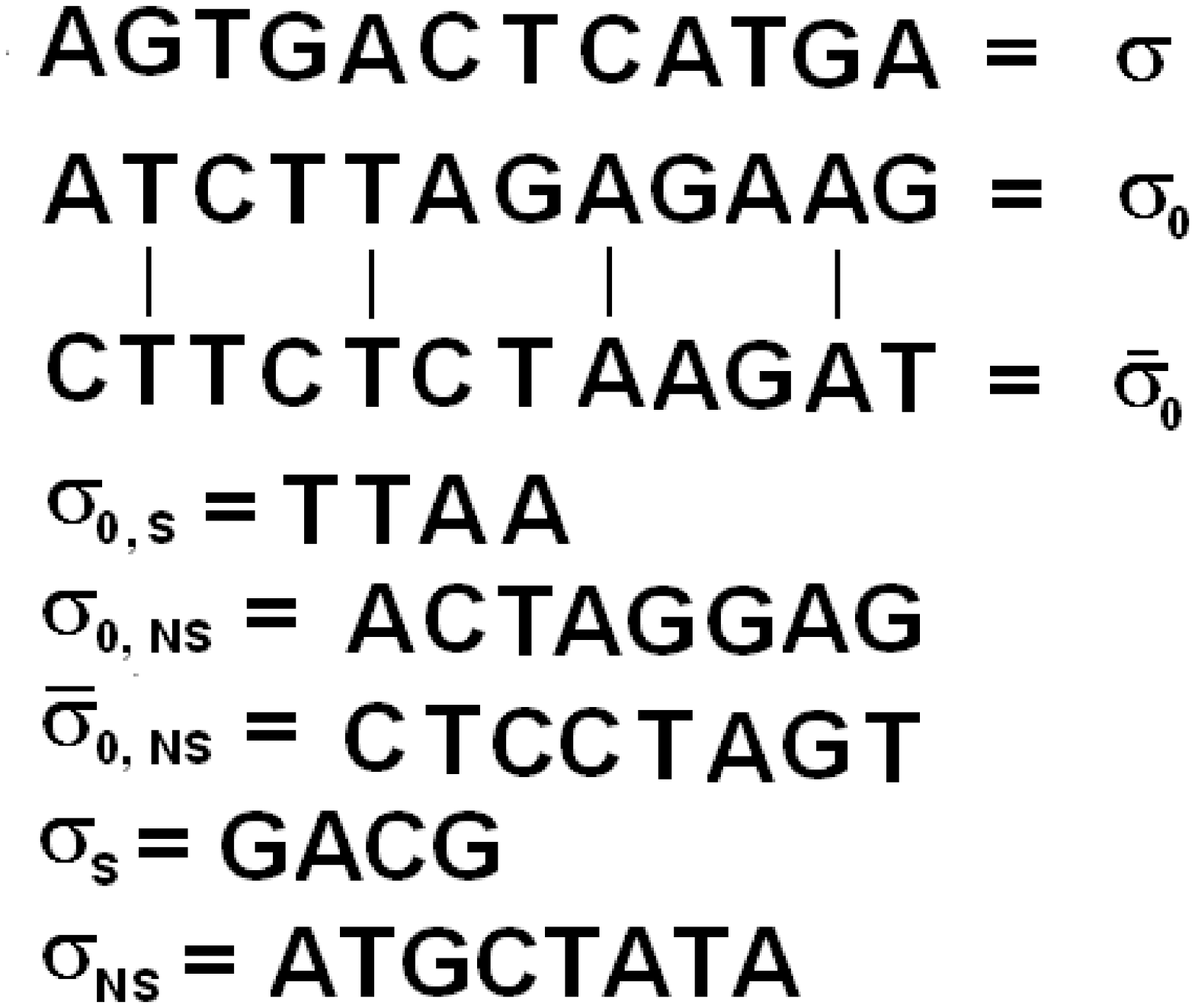}
\caption{Illustration of sequence decomposition into $ \sigma_S $ and
$ \sigma_{NS} $ components.}
\end{figure}

Given some gene sequence $ \sigma $, we can then characterize it by the 
following numbers:  (1) $ l_S \equiv HD(\sigma_S, \sigma_{0, S}) $.
(2) $ l_{NS} \equiv HD(\sigma_{NS}, \sigma_{0, NS}) $.  (3)  $ \bar{l}_{NS} 
\equiv HD(\sigma_{NS}, \bar{\sigma}_{0, NS}) $.  (4)  $ \tilde{l}_{NS} \equiv
l_{NS} + \bar{l}_{NS} - l_I $, so $ \tilde{l}_{NS} $ is simply the number
of positions where $ \sigma_{NS} $ differs from both $ \sigma_{0, NS} $ and
$ \bar{\sigma}_{0, NS} $.

Now, any $ \sigma' $ may be generated from any $ \sigma $ by making the
appropriate base changes.  We can make changes to $ \sigma $ as follows:
\begin{enumerate}
\item Let $ l_{1, S} $ denote the number of changes to $ \sigma_{S} $
where $ \sigma_{S} $ and $ \sigma_{0,S} $ are identical.  There are
$ {{L-l_I-l_S}\choose {l_{1,S}}} (2S-1)^{l_{1,S}} $ possibilities for this
set of changes.
\item Let $ l_{2, S} $ denote the number of changes to $ \sigma_{S} $
back to the corresponding base in $ \sigma_{0,S} $, where $ \sigma_{S} $ 
and $ \sigma_{0,S} $ are distinct.  The degeneracy in this case is
$ {{l_S} \choose {l_{2,S}}} $.
\item Let $ l_{3, S} $ denote the number of changes to $ \sigma_{S} $
to bases distinct from the corresponding bases in $ \sigma_{0,S} $, where
$ \sigma_{S} $ and $ \sigma_{0,S} $ are distinct.  The degeneracy is
$ {{l_S - l_{2,S}} \choose l_{3,S}} (2S-2)^{l_{3,S}} $.
\item Let $ l_{11, NS} $ denote the number of changes to $ \sigma_{NS} $
where $ \sigma_{NS} $, $ \sigma_{0, NS} $ are identical, to bases other
than the corresponding ones in $ \bar{\sigma}_{0, NS} $.  The degeneracy is
$ {{l_I - l_{NS}} \choose l_{11, NS}} (2S-2)^{l_{11,NS}} $.
\item Let $ l_{12, NS} $ denote the number of changes to $ \sigma_{NS} $
where $ \sigma_{NS} $, $ \sigma_{0, NS} $ are identical, to the corresponding
bases in $ \bar{\sigma}_{0, NS} $.  The degeneracy is $ {{l_I - l_{NS} -
l_{11, NS}} \choose l_{12, NS}} $.
\item Let $ \bar{l}_{11, NS} $ denote the number of changes to $ \sigma_{NS} $
where $ \sigma_{NS} $, $ \bar{\sigma}_{0, NS} $ are identical, to bases other
than the corresponding ones in $ \sigma_{0, NS} $.  The degeneracy is
$ {{l_I - \bar{l}_{NS}} \choose \bar{l}_{11, NS}} (2S-2)^{\bar{l}_{11, NS}} $.
\item Let $ \bar{l}_{12, NS} $ denote the number of changes to $ \sigma_{NS} $
where $ \sigma_{NS} $, $ \bar{\sigma}_{0, NS} $ are identical, to the
corresponding bases in $ \sigma_{0, NS} $.  The degeneracy is
$ {{l_I - \bar{l}_{NS} - \bar{l}_{11, NS}} \choose \bar{l}_{12, NS}} $.
\item Let $ l_{2, NS} $ denote the number of changes to $ \sigma_{NS} $, where
$ \sigma_{NS} $ is distinct from $ \sigma_{0, NS} $ and $ \bar{\sigma}_{0, NS} 
$, and the bases are changed to the corresponding bases in $ \sigma_{NS, 0} $.
The degeneracy is $ {{\tilde{l}_{NS}} \choose {l_{2, NS}}} $.
\item Let $ \bar{l}_{2, NS} $ denote the number of changes to $ \sigma_{NS} $,
where $ \sigma_{NS} $ is distinct from $ \sigma_{0, NS} $ and 
$ \bar{\sigma}_{0, NS} $, and the bases are changed to the corresponding bases
in $ \bar{\sigma}_{NS, 0} $.  The degeneracy is $ {{\tilde{l}_{NS} - l_{2, NS}}
\choose \tilde{l}_{2, NS}} $.
\item Let $ l_{3, NS} $ denote the number of changes to $ \sigma_{NS} $,
where $ \sigma_{NS} $ is distinct from $ \sigma_{0, NS} $ and
$ \bar{\sigma}_{0, NS} $, and the bases are changed to bases other than
the corresponding ones in $ \sigma_{NS, 0} $ and $ \bar{\sigma}_{NS, 0} $.
The degeneracy is $ {{\tilde{l}_{NS} - l_{2, NS} - \tilde{l}_{2, NS}} \choose
l_{3, NS}} (2S-3)^{l_{3, NS}} $. 
\end{enumerate}

The series of changes to $ \sigma $ defined above yield a $ \sigma' $ which
is at a Hamming distance of $ l_{1, S} + l_{2, S} + l_{3, S} + l_{11, NS}
+ l_{12, NS} + \bar{l}_{11, NS} + \bar{l}_{12, NS} + l_{2, NS} + 
\bar{l}_{2, NS} + l_{3, NS} $ from $ \sigma $.  Furthermore, the values of
$ l_S $, $ l_{NS} $, $ \bar{l}_{NS} $, and $ \tilde{l}_{NS} $ for $ \sigma' $
are given by,
\begin{eqnarray}
&   & l_{S}' = l_S + l_{1, S} - l_{2, S} \nonumber \\
&   & l_{NS}' = l_{NS} + l_{11, NS} + l_{12, NS} - \bar{l}_{12, NS} - 
                l_{2, NS} \nonumber \\
&   & \bar{l}_{NS}' = \bar{l}_{NS} + \bar{l}_{11, NS} + \bar{l}_{12, NS}  -
                      l_{12, NS} - \bar{l}_{2, NS} \nonumber \\
&   & \tilde{l}_{NS}' = \tilde{l}_{NS} + l_{11, NS} + \bar{l}_{11, NS} - 
                        l_{2, NS} - \bar{l}_{2, NS}
\end{eqnarray}

Now, we will assume that $ y_{\sigma} $ depends only on $ l_S $, $ l_{NS} $,
and $ \bar{l}_{NS} $.  At time $ t = 0 $, we start the evolution by setting
$ y_{\sigma} = 0 $ for all $ \sigma \neq \sigma_0, \bar{\sigma}_0 $, and
$ y_{\sigma_0} = y_{\bar{\sigma}_0} = 1/2 $ if $ \sigma_0 \neq \bar{\sigma}_0 
$, and $ 1 $ if $ \sigma_0 = \bar{\sigma}_0 $.  Therefore, $ y_{\sigma} = 0 $
unless $ l_S = 0 $ and $ l_{NS} = 0 $ or $ \bar{l}_{NS} = 0 $.  So certainly
$ y_{\sigma} $ depends only on $ l_S $, $ l_{NS} $, and $ \bar{l}_{NS} $ at
the start of the evolution.  Also, by similar reasoning, we see that the
fitness landscape $ \{\kappa_{\sigma}\} $ also depends only on $ l_S $,
$ l_{NS} $, $ \bar{l}_{NS} $.  If we can show that this implies that 
$ d y_{\sigma}/dt $ depends only on $ l_S $, $ l_{NS} $, $ \bar{l}_{NS} $,
then $ y_{\sigma} $ depends only on $ l_S $, $ l_{NS} $, $ \bar{l}_{NS} $
throughout the evolution.

So, consider some time $ t $ for which the $ y_{\sigma} $ depend only on 
$ l_S $, $ l_{NS} $, $ \bar{l}_{NS} $, for all given $ \sigma $ characterized 
by $ l_S $, $ l_{NS} $, $ \bar{l}_{NS} $.  Then we may write $ y_{l_S, l_{NS},
\bar{l}_{NS}} \equiv y_{\sigma} $.  We also write $ \kappa_{\sigma} =
\kappa_{l_S, l_{NS}, \bar{l}_{NS}} $.  Summing over the contributions from
the $ \sigma' $, obtained by the base changes described above, we obtain,
\begin{widetext}
\begin{eqnarray}
\frac{d y_{l_S, l_{NS}, \bar{l}_{NS}}}{dt} 
& = &
2
\sum_{l_{1, S} = 0}^{L - l_I - l_S} \sum_{l_{2, S} = 0}^{l_S}
\sum_{l_{3, S} = 0}^{l_S - l_{2, S}} \sum_{l_{11, NS} = 0}^{l_I - l_{NS}}
\sum_{l_{12, NS} = 0}^{l_I - l_{NS} - l_{11, NS}}
\sum_{\bar{l}_{11, NS} = 0}^{l_I - \bar{l}_{NS}}
\sum_{\bar{l}_{12, NS} = 0}^{l_I - \bar{l}_{NS} - \bar{l}_{11, NS}}
\sum_{l_{2, NS} = 0}^{\tilde{l}_{NS}}
\sum_{\bar{l}_{2, NS} = 0}^{\tilde{l}_{NS} - l_{2, NS}}
\sum_{l_{3, NS} = 0}^{\tilde{l}_{NS} - l_{2, NS} - \bar{l}_{2, NS}} 
\nonumber \\
&   &
{{L - l_I - l_S}\choose {l_{1, S}}} {l_S \choose l_{2, S}}
{{l_S - l_{2, S}} \choose l_{3, S}}{{l_I - l_{NS}} \choose l_{11, NS}}
{{l_I - l_{NS} - l_{11, NS}} \choose l_{12, NS}} 
{{l_I - \bar{l}_{NS}} \choose \bar{l}_{11, NS}} \times \nonumber \\
&   & 
{{l_I - \bar{l}_{NS} - \bar{l}_{11, NS}} \choose \bar{l}_{12, NS}}
{\tilde{l}_{NS} \choose l_{2, NS}} {{\tilde{l}_{NS} - l_{2, NS}} \choose
\bar{l}_{2, NS}} {{\tilde{l}_{NS} - l_{2, NS} - \bar{l}_{2, NS}} \choose
l_{3, NS}} \times \nonumber \\
&   &
(2S-1)^{l_{1,S}} (2S-2)^{l_{3,S}} (2S-2)^{l_{11, NS}} (2S-2)^{\bar{l}_{11, NS}}
(2S-3)^{l_{3, NS}} \times \nonumber \\
&   &
(\frac{\eps/2}{2S-1})^{l_{1, S} + l_{2, S} + l_{3, S} + l_{11, NS} + 
\bar{l}_{11, NS} + l_{12, NS} + \bar{l}_{12, NS} + l_{2, NS} +
\bar{l}_{2, NS} + l_{3, NS}} \times \nonumber \\
&   &
(1 - \frac{\eps}{2})^{L - l_{1, S} - l_{2, S}
- l_{3, S} - l_{11, NS} - \bar{l}_{11, NS} - l_{12, NS} - \bar{l}_{12, NS}
- l_{2, NS} - \bar{l}_{2, NS} - l_{3, NS}} \times \nonumber \\
&   &
\kappa_{l_S + l_{1, S} - l_{2, S}, l_{NS} + l_{11, NS} + l_{12, NS} -
\bar{l}_{12, NS} - l_{2, NS}, \bar{l}_{NS} + \bar{l}_{11, NS} +
\bar{l}_{12, NS} - l_{12, NS} - \bar{l}_{2, NS}} \times \nonumber \\
&   & 
y_{l_S + l_{1, S} - l_{2, S}, l_{NS} + l_{11, NS} + l_{12, NS} -
\bar{l}_{12, NS} - l_{2, NS}, \bar{l}_{NS} + \bar{l}_{11, NS} +
\bar{l}_{12, NS} - l_{12, NS} - \bar{l}_{2, NS}} \nonumber \\
&   &
- (\kappa_{l_S, l_{NS}, \bar{l}_{NS}} + \bar{\kappa}(t)) y_{l_S, l_{NS},
\bar{l}_{NS}}
\end{eqnarray}
Note from the sum that $ dy_{\sigma}/dt = dy_{\sigma'}/dt $ for any two
$ \sigma $, $ \sigma' $ characterized by the same $ l_S $, $ l_{NS} $, and
$ \bar{l}_{NS} $.  Therefore, the assumption that $ y_{\sigma} $ is determined
by $ l_S $, $ l_{NS} $, $ \bar{l}_{NS} $ is justified.

We may sum over $ l_{3, S} $ and $ l_{3, NS} $ to obtain,
\begin{eqnarray}
\frac{d y_{l_S, l_{NS}, \bar{l}_{NS}}}{dt} 
& = &
2
\sum_{l_{1, S} = 0}^{L - l_I - l_S} \sum_{l_{2, S} = 0}^{l_S}
\sum_{l_{11, NS} = 0}^{l_I - l_{NS}}
\sum_{l_{12, NS} = 0}^{l_I - l_{NS} - l_{11, NS}}
\sum_{\bar{l}_{11, NS} = 0}^{l_I - \bar{l}_{NS}}
\sum_{\bar{l}_{12, NS} = 0}^{l_I - \bar{l}_{NS} - \bar{l}_{11, NS}}
\sum_{l_{2, NS} = 0}^{\tilde{l}_{NS}}
\sum_{\bar{l}_{2, NS} = 0}^{\tilde{l}_{NS} - l_{2, NS}} \nonumber \\
&   &
{{L - l_I - l_S} \choose l_{1, S}}{l_S \choose l_{2, S}}{{l_I - l_{NS}}
\choose l_{11, NS}}{{l_I - l_{NS} - l_{11, NS}} \choose l_{12, NS}}
\times \nonumber \\
&   &
{{l_I - \bar{l}_{NS}} \choose \bar{l}_{11, NS}} {{l_I - \bar{l}_{NS} -
\bar{l}_{11, NS}} \choose \bar{l}_{12, NS}}{\tilde{l}_{NS} \choose l_{2, NS}}
{{\tilde{l}_{NS} - l_{2, NS}} \choose \bar{l}_{2, NS}} \times \nonumber \\
&   &
(2S-2)^{l_{11, NS} + \bar{l}_{11, NS}} (\frac{\eps}{2})^{l_{1, S}}
(\frac{\eps/2}{2S-1})^{l_{2, S} + l_{11, NS} + \bar{l}_{11, NS} + l_{12, NS}
+ \bar{l}_{12, NS} + l_{2, NS} + \bar{l}_{2, NS}} \times \nonumber \\
&   & 
(1 - \frac{\eps/2}{2S-1})^{l_S - l_{2, S}} (1 - \frac{\eps}{2S-1})^
{\tilde{l}_{NS} - l_{2, NS} - \bar{l}_{2, NS}} \times \nonumber \\
&   &
(1 - \frac{\eps}{2})^{L - l_S - \tilde{l}_{NS} - l_{1, S} - l_{11, NS} -
\bar{l}_{11, NS} - l_{12, NS} - \bar{l}_{12, NS}} \times \nonumber \\
&   &
\kappa_{l_S + l_{1, S} - l_{2, S}, l_{NS} + l_{11, NS} + l_{12, NS} -
\bar{l}_{12, NS} - l_{2, NS}, \bar{l}_{NS} + \bar{l}_{11, NS} +
\bar{l}_{12, NS} - l_{12, NS} - \bar{l}_{2, NS}} \times \nonumber \\
&   &
y_{l_S + l_{1, S} - l_{2, S}, l_{NS} + l_{11, NS} + l_{12, NS} - 
\bar{l}_{12, NS} - l_{2, NS}, \bar{l}_{NS} + \bar{l}_{11, NS} +
\bar{l}_{12, NS} - l_{12, NS} - \bar{l}_{2, NS}} - \nonumber \\
&   &
(\kappa_{l_S, l_{NS}, \bar{l}_{NS}} + \bar{\kappa}(t)) y_{l_S, l_{NS}, 
\bar{l}_{NS}}
\end{eqnarray}

Now, the total number of sequences $ \sigma $ characterized by the Hamming
distances $ l_S $, $ l_{NS} $, and $ \bar{l}_{NS} $ is given by
$ C_{l_S, l_{NS}, \bar{l}_{NS}} = {{L - l_I} \choose l_S}{l_I \choose l_{NS}}
{l_{NS} \choose \tilde{l}_{NS}} (2S-1)^{l_S} (2S-2)^{\tilde{l}_{NS}} $.
Then define $ z_{l_S, l_{NS}, \bar{l}_{NS}} = C_{l_S, l_{NS}, \bar{l}_{NS}}
y_{l_S, l_{NS}, \bar{l}_{NS}} $.  We may convert our differential equations
from the $ y $ to the $ z $ representations.  After some tedious algrebra,
the final result is,
\begin{eqnarray}
\frac{d z_{l_S, l_{NS}, \bar{l}_{NS}}}{dt} 
& = &
2
\sum_{l_{1, S} = 0}^{L - l_I - l_S} \sum_{l_{2, S} = 0}^{l_S}
\sum_{l_{11, NS} = 0}^{l_I - l_{NS}}
\sum_{l_{12, NS} = 0}^{l_I - l_{NS} - l_{11, NS}}
\sum_{\bar{l}_{11, NS} = 0}^{l_I - \bar{l}_{NS}}
\sum_{\bar{l}_{12, NS} = 0}^{l_I - \bar{l}_{NS} - \bar{l}_{11, NS}}
\sum_{l_{2, NS} = 0}^{\tilde{l}_{NS}}
\sum_{\bar{l}_{2, NS} = 0}^{\tilde{l}_{NS} - l_{2, NS}} \nonumber \\
&   &
{{l_{1, S} + l_S - l_{2, S}} \choose l_{1, S}} (\frac{\eps/2}{2S-1})^{l_{1, S}}
(1 - \frac{\eps/2}{2S-1})^{l_S - l_{2, S}} \times \nonumber \\
&   &
{{L - l_I - l_S - l_{1, S} + l_{2, S}} \choose l_{2, S}} (\frac{\eps}{2})^
{l_{2, S}} \times \nonumber \\
&   &
{{\tilde{l}_{NS} - l_{2, NS} - \bar{l}_{2, NS} + l_{11, NS} + \bar{l}_{11, NS}}
\choose \bar{l}_{11, NS}}{{\tilde{l}_{NS} - l_{2, NS} - \bar{l}_{2, NS} + 
l_{11, NS}} \choose l_{11, NS}} \times \nonumber \\
&   &
(\frac{\eps/2}{2S-1})^{l_{11, NS} + \bar{l}_{11, NS}} 
(1 - \frac{\eps}{2S-1})^{\tilde{l}_{NS} - l_{2, NS} - \bar{l}_{2, NS}} 
\times \nonumber \\
&   &
{{l_I - l_{NS} - l_{11, NS} - l_{12, NS} + l_{2, NS} + \bar{l}_{12, NS}} 
\choose \bar{l}_{12, NS}} {{l_I - l_{NS} - l_{11, NS} - l_{12, NS} + l_{2, NS}}
\choose l_{2, NS}} \times \nonumber \\
&   &
(\frac{\eps/2}{2S-1})^{\bar{l}_{12, NS} + l_{2, NS}} \times \nonumber \\
&   &
{{l_I - \bar{l}_{NS} - \bar{l}_{11, NS} - \bar{l}_{12, NS} + l_{12, NS} +
\bar{l}_{2, NS}} \choose \bar{l}_{2, NS}}
{{l_I - \bar{l}_{NS} - \bar{l}_{11, NS} - \bar{l}_{12, NS} + l_{12, NS}} 
\choose l_{12, NS}} \times \nonumber \\
&   &
(\frac{\eps/2}{2S-1})^{\bar{l}_{2, NS} + l_{12, NS}}
(2S-2)^{l_{2, NS} + \bar{l}_{2, NS}}
(1 - \frac{\eps}{2})^{L - l_S - \tilde{l}_{NS} - l_{1, S} - l_{11, NS} -
\bar{l}_{11, NS} - l_{12, NS} - \bar{l}_{12, NS}} \times \nonumber \\ 
&   &
\kappa_{l_S + l_{1, S} - l_{2, S}, l_{NS} + l_{11, NS} + l_{12, NS} -
\bar{l}_{12, NS} - l_{2, NS}, \bar{l}_{NS} + \bar{l}_{11, NS} +
\bar{l}_{12, NS} - l_{12, NS} - \bar{l}_{2, NS}} \times \nonumber \\
&   &
z_{l_S + l_{1, S} - l_{2, S}, l_{NS} + l_{11, NS} + l_{12, NS} -
\bar{l}_{12, NS} - l_{2, NS}, \bar{l}_{NS} + \bar{l}_{11, NS} +
\bar{l}_{12, NS} - l_{12, NS} - \bar{l}_{2, NS}} - \nonumber \\
&   &
(\kappa_{l_S, l_{NS}, \bar{l}_{NS}} + \bar{\kappa}(t)) z_{l_S, l_{NS}, 
\bar{l}_{NS}}
\end{eqnarray}
\end{widetext}

\subsection{The Infinite Sequence Length Equations}

We are now in a position to derive the infinite sequence length form
of the quasispecies equations.  We allow $ L \rightarrow \infty $ while
keeping $ \mu \equiv L \eps $ fixed.  Furthermore, let us define $ f_I =
l_I/L $, so $ f_I $ is the fraction of bases in $ \sigma_0 $ and 
$ \bar{\sigma}_0 $ which differ.  If we let $ p(f_I) $ denote the probability
density for $ f_I $, then in the limit of infinite sequence length we
obtain that $ p(f_I) \rightarrow \delta(f_I - (1 - \frac{1}{2S})) $,
where $ \delta $ is the Dirac $ \delta $-function.  Therefore, we take $ f_I
= 1 - \frac{1}{2S} $ in the $ L \rightarrow \infty $ limit.

A slight complication arises in the infinite sequence limit, namely, that
$ l_I = f_I L \rightarrow \infty $ as $ L \rightarrow \infty $.  This means 
that it is impossible for $ l_{NS} $ and $ \bar{l}_{NS} $ to simultaneously 
be finite.  For if $ l_{NS} $ is finite, then $ \bar{l}_{NS} = l_I - l_{NS} + 
\tilde{l}_{NS} = \infty $ and vice versa.  The appropriate way to solve these 
equations is therefore to solve for finite values of $ l_S $, $ l_{NS} $, and 
$ \tilde{l}_{NS} $.  Then we can redenote $ z_{l_S, l_{NS}, \bar{l}_{NS}} $
by $ z_{l_S, l_{NS}, \tilde{l}_{NS}} $, and solve in the infinite sequence
limit.  The symmetry of the landscape allows us to obtain the finite
$ \bar{l}_{NS} $ population fractions as well, since the population fraction
for finite $ l_S $, $ \bar{l}_{NS} $, and $ \tilde{l}_{NS} $ is then simply
given by $ z_{l_S, \bar{l}_{NS}, \tilde{l}_{NS}} $.

In the following subsection, we show that as $ L \rightarrow \infty $, the 
only terms which survive the limiting process are the $ l_{1, S} = 
l_{11, NS} = \bar{l}_{11, NS} = \bar{l}_{2, NS} = l_{12, NS} = 0 $ terms.  
We also have,
\begin{widetext}
\begin{equation}
{{L - l_I - l_S + l_{2, S}} \choose l_{2, S}} (\frac{\eps}{2})^{l_{2, S}}
\rightarrow \frac{1}{l_{2, S}!}(\frac{(1 - f_I) \mu}{2})^{l_{2, S}}
\end{equation}  
\begin{equation}
{{l_I - l_{NS} + l_{2, NS} + \bar{l}_{12, NS}} \choose \bar{l}_{12, NS}}
{{l_I - l_{NS} + l_{2, NS}} \choose l_{2, NS}} (\frac{\eps/2}{2S-1})^{
\bar{l}_{12, NS} + l_{2, NS}} \rightarrow 
\frac{1}{l_{2, NS}! \bar{l}_{12, NS}!} (\frac{f_I \mu}{2 (2S-1)})^
{\bar{l}_{12, NS} + l_{2, NS}}
\end{equation}
\begin{equation}
(1 - \frac{\eps}{2})^{L - l_S - \tilde{l}_{NS} - l_{1, S} - l_{11, NS} 
- \bar{l}_{11, NS} - l_{12, NS} - \bar{l}_{12, NS}} \rightarrow e^{-\mu/2}
\end{equation}
Using the fact that $ f_I \rightarrow 1 - \frac{1}{2S} $ as $ L \rightarrow
\infty $, we obtain, after some manipulation (and after redenoting
$ \kappa_{l_S, l_{NS}, \bar{l}_{NS}} $ by $ \kappa_{l_S, l_{NS}, 
\tilde{l}_{NS}} $), the infinite sequence length equations,
\begin{eqnarray}
\frac{d z_{l_S, l_{NS}, \tilde{l}_{NS}}}{dt} 
& = &
2 e^{-\frac{\mu}{2}} \sum_{l_{1, S} = 0}^{l_S} \sum_{l_{1, NS} = 0}^{l_{NS} -
\tilde{l}_{NS}} \sum_{l_{2, NS} = 0}^{\tilde{l}_{NS}} 
\frac{1}{l_{1, S}! l_{1, NS}! l_{2, NS}!} (\frac{\mu}{4S})^{l_{1, S} +
l_{1, NS} + l_{2, NS}} (2S-2)^{l_{2, NS}} \times \nonumber \\
&   &
\kappa_{l_S - l_{1, S}, l_{NS} - l_{1, NS} - l_{2, NS}, \tilde{l}_{NS} -
l_{2, NS}}
z_{l_S - l_{1, S}, l_{NS} - l_{1, NS} - l_{2, NS}, \tilde{l}_{NS} - l_{2, NS}}
- \nonumber \\
&   &
(\kappa_{l_S, l_{NS}, \tilde{l}_{NS}} + \bar{\kappa}(t)) z_{l_S, l_{NS},
\tilde{l}_{NS}}
\end{eqnarray}
where we have redenoted $ l_{2,S} $ by $ l_{1, S} $, and $ \bar{l}_{12,NS} $
by $ l_{1,NS} $.

It should be clear that $ z_{0, 0, 0} = y_{\sigma_0} $.  Therefore, 
$ 2 z_{0, 0, 0} $ is the total fraction of the population with genome
$ \{\sigma_0, \bar{\sigma}_0 \} $.  This gives, $ \bar{\kappa}(t) =
2(k-1) z_{0, 0, 0} + 1 $.
\end{widetext}

Now, as $ L \rightarrow \infty $, the sequences $ \sigma_0 $ and 
$ \bar{\sigma}_0 $ become infinitely separated.  Therefore, we expect
that the values of $ z_{l_S, l_{NS}, \tilde{l}_{NS}} $ for finite
$ l_S $, $ l_{NS} $, $ \tilde{l}_{NS} $ to be dictated by the single
fitness peak at $ \sigma_0 $.  Thus, for large $ L $, we expect to
obtain a locally single fitness peak model in which we can then assume
that $ y_{\sigma} $ depends only on the Hamming distance $ l_S + l_{NS} $
to $ \sigma_0 $.  In the following subsection, we prove this rigorously.  We 
may then group the population into Hamming classes, as with the single 
fitness peak for conservative replication.  Specifically, we define $ w_l = 
\sum_{l_{NS} = 0}^{l} \sum_{\tilde{l}_{NS} = 0}^{l_{NS}} z_{l - l_{NS},
l_{NS}, \tilde{l}_{NS}} $, and finally obtain the infinite sequence
length equations given by Eq. (17).

\subsection{Additional Calculational Details}

\subsubsection{Derivation of the Infinite Sequence Length Equations from the
Finite Sequence Length Equatiions}

In this appendix, we derive the infinite sequence length form
for $ dz_{l_S, l_{NS}, \tilde{l}_{NS}}/dt $ from the corresponding
finite sequence length equations.  Before proceeding, however, we derive
some basic inequalities which we will need to use.  First of all,
note that each $ z_{l_S, l_{NS}, \tilde{l}_{NS}} $ must be $ \leq 1 $.
Furthermore, note that $ \kappa_{l_S, l_{NS}, \tilde{l}_{NS}} \leq k $.
We also have, $ {{m + n} \choose m} \lambda^m = \prod_{i = 1}^{m}
{\frac{n+i}{i} \lambda} \leq ((n+1)\lambda)^m $, and
$ (1 - \lambda)^n \leq 1 $ for $ \lambda \in [0, 1] $.

We wish to show that in the limit of $ L \rightarrow \infty $, the
only terms which contribute to the dynamical equations are the
$ l_{1, S} = l_{11,NS} = \bar{l}_{11, NS} = \bar{l}_{2, NS} = l_{12,NS} = 0
$ terms.  We prove this by showing that for each of the above indices, 
the total contribution from all the nonzero terms becomes arbitrarily 
small as $ L \rightarrow \infty $ with $ \mu = L \eps $ held fixed.

So, we start with the $ l_{1, S} $ index.  From the inequalities given
above, we may note that the summand of Eq. (A4), denoted by $ S_{l_S, l_{NS}, 
\tilde{l}_{NS}, l_{1,S}, l_{2,S}, l_{11,NS}, \bar{l}_{11,NS}, 
l_{12,NS}, \bar{l}_{12, NS}, l_{2, NS}, \bar{l}_{2, NS}} $, has the upper 
bound,
\begin{widetext}
\begin{eqnarray}
&   &
S_{l_S, l_{NS}, \tilde{l}_{NS}, l_{1,S}, l_{2,S}, l_{11,NS}, \bar{l}_{11,NS}, 
l_{12, NS}, \bar{l}_{12, NS}, l_{2, NS}, \bar{l}_{2, NS}} \nonumber \\
&   &
\leq 
k ((l_S + 1) (\frac{\eps/2}{2S-1}))^{l_{1, S}}
((L + 1) \frac{\eps}{2})^{l_{2, S}}
((\tilde{l}_{NS} + 1)(\frac{\eps/2}{2S-1}))^{l_{11,NS}}
((\tilde{l}_{NS} + 1)(\frac{\eps/2}{2S-1})^{\bar{l}_{11,NS}} 
\times \nonumber \\ 
&    &
((l_I + 1)(\frac{\eps/2}{2S-1}))^{\bar{l}_{12,NS}}
((l_I + 1)(\frac{2S-2}{2S-1}\frac{\eps}{2}))^{l_{2,NS}}
((l_{NS} - \tilde{l}_{NS} + 1) (\frac{2S-2}{2S-1} \frac{\eps}{2})^
{\bar{l}_{2,NS}}
((l_{NS} - \tilde{l}_{NS} + 1) (\frac{\eps/2}{2S-1})^{l_{12,NS}} \nonumber \\
\end{eqnarray}
Now, at fixed $ \mu $, choose $ L $ to be sufficiently large so that
$ (L + 1) \frac{\eps}{2} = \frac{1}{2} (\mu + \eps) < \mu $.  Then
certainly $ (l_I + 1) (\frac{\eps/2}{2S-1}) < \frac{\mu}{2S-1} $.
We then have,
\begin{eqnarray}
&   &
\sum_{l_{1,S} = 1}^{L - l_I - l_S}\sum_{l_{2,S} = 0}^{l_S} \sum_{l_{11,NS} = 0}
^{l_I - l_{NS}} \sum_{l_{12, NS} = 0}^{l_I - l_{NS} - l_{11,NS}}
\sum_{\bar{l}_{11,NS} = 0}^{l_{NS} - \tilde{l}_{NS}}
\sum_{\bar{l}_{12,NS} = 0}^{l_{NS} - \tilde{l}_{NS} - \bar{l}_{11,NS}}
\sum_{l_{2, NS} = 0}^{\tilde{l}_{NS}} \sum_{\bar{l}_{2,NS} = 0}^
{\tilde{l}_{NS} - l_{2,NS}} \nonumber \\
&   &
{S_{l_S, l_{NS}, \tilde{l}_{NS}, l_{1,S}, l_{2,S}, l_{11,NS}, l_{12,NS},
\bar{l}_{11,NS}, \bar{l}_{12,NS}, l_{2,NS}, \bar{l}_{2,NS}}} \nonumber \\
&   &
\leq 
k \sum_{l_{1, S} = 1}^{L-l_I-l_S}{((l_S + 1) (\frac{\eps/2}{2S-1}))^
{l_{1, S}}}
\sum_{l_{2, S} = 0}^{l_S}{\mu^{l_{2,S}}}
\sum_{l_{11,NS} = 0}^{l_I - l_{NS}}
{((\tilde{l}_{NS} + 1)(\frac{\eps/2}{2S-1}))^{l_{11,NS}}} \times \nonumber \\
&      &
\sum_{l_{12,NS} = 0}^{l_I - l_{NS}}
{((l_{NS} - \tilde{l}_{NS} + 1) (\frac{\eps/2}{2S-1}))^{l_{12,NS}}} 
\sum_{\bar{l}_{11, NS} = 0}^{l_{NS} - \tilde{l}_{NS}}
{((\tilde{l}_{NS} + 1) (\frac{\eps/2}{2S-1}))^{\bar{l}_{11,NS}}} 
\times \nonumber \\
&      &
\sum_{\bar{l}_{12,NS} = 0}^{l_{NS} - \tilde{l}_{NS}}
{(\frac{\mu}{2S-1})^{\bar{l}_{12,NS}}}
\sum_{l_{2,NS} = 0}^{\tilde{l}_{NS}}
{\mu^{l_{2,NS}}}
\sum_{\bar{l}_{2, NS} = 0}^{\tilde{l}_{NS}}
{((l_{NS} - \tilde{l}_{NS} + 1) (\frac{\eps}{2}))^{\bar{l}_{2,NS}}}
\end{eqnarray}
\end{widetext}
Now, let $ A_l \equiv \sum_{k = 0}^{l}{\mu^k} $.  Also, note
that $ \sum_{n = m}^{\infty}{\lambda^n} = \frac{\lambda^m}{1 - \lambda} $,
for $ |\lambda| < 1 $.  Therefore, note that an upper bound for the product 
given above is simply, $ k A_{l_S + l_{NS}}^3 ((l_S + l_{NS} + 1) 
(\eps/2))/(1 - (l_S + l_{NS} + 1)(\eps/2))^{5} $.  Therefore, 
as $ L \rightarrow \infty $, so that $ \eps \rightarrow 0 $ in such a way 
that $ \mu $ is fixed, we see that the contribution of the $ l_{1, S} > 0 $ 
terms to the evolution dynamics approaches $ 0 $.  Therefore, we need only 
consider the $ l_{1, S} = 0 $ terms in the limit of infinite sequence length.  
Using a similar argument to the one given above, we can systematically 
eliminate the contributions from the $ l_{11,NS}, l_{12,NS}, \bar{l}_{11,NS},
\bar{l}_{2,NS} > 0 $ as well.  This establishes the infinite sequence
length form of our differential equations.  We should note that convergence
to the infinite sequence length form is not uniform, as can be seen by
the $ l_S + l_{NS} $ dependence of our upper bound.

\subsubsection{Simplification of the Infinite Sequence Length Equations}

We wish to show that, as $ L \rightarrow \infty $, we may assume that 
$ y_{l_S, l_{NS}, \tilde{l}_{NS}} $ becomes dependent only on 
$ l_S + l_{NS} $, which will thereby allow us to considerably simplify
the infinite sequence length equations (Eq. (A4)). 

To proceed with this simplification, let us first determine the effect
that $ y_{l_S, l_{NS}, \tilde{l}_{NS}} $ depending only on $ l_S + l_{NS} $
has on $ z_{l_S, l_{NS}, \tilde{l}_{NS}} $.  We have, $ z_{l_S, l_{NS}, 
\tilde{l}_{NS}} = C_{l_S, l_{NS}, \tilde{l}_{NS}} 
y_{l_S, l_{NS}, \tilde{l}_{NS}} $.  But, $ y_{l_S, l_{NS}, \tilde{l}_{NS}} =
y_{l_S, l_{NS}, 0} = z_{l_S, l_{NS}, 0}/C_{l_S, l_{NS}, 0} $.  Putting 
everything together, we obtain,
\begin{equation}
z_{l_S, l_{NS}, \tilde{l}_{NS}} = {l_{NS} \choose \tilde{l}_{NS}}
(2S-2)^{\tilde{l}_{NS}} z_{l_S, l_{NS}, 0}
\end{equation}
A similar procedure yields,
\begin{eqnarray}
z_{l_S, l_{NS}, 0} 
& = &
{{l_S + l_{NS}} \choose l_{NS}} \frac{l_I! (L - l_I - l_S - l_{NS})!}
{(l_I - l_{NS})! (L - l_I - l_S)!} \times \nonumber \\
&   & 
(2S-1)^{-l_{NS}} z_{l_S + l_{NS}, 0, 0}
\end{eqnarray}
As $ L, l_I \rightarrow \infty $, we get,
\begin{eqnarray}
&   & 
\frac{l_I!}{(l_I - l_{NS})!}\frac{(L - l_I - l_S - l_{NS})!}
{(L - l_I - l_S)!} \rightarrow \nonumber \\ 
&   &
(\frac{l_I}{L - l_I})^{l_{NS}} = (\frac{f_I}{1 - f_I})^{l_{NS}}
= (2S-1)^{l_{NS}}
\end{eqnarray}
giving, $ z_{l_S, l_{NS}, 0} = {{l_S + l_{NS}} \choose l_{NS}}
z_{l_S + l_{NS}, 0, 0} $.  Therefore,
\begin{equation}
z_{l_S, l_{NS}, \tilde{l}_{NS}} = {{l_S + l_{NS}} \choose l_{NS}}
{l_{NS} \choose \tilde{l}_{NS}} (2S-2)^{\tilde{l}_{NS}} z_{l_S + l_{NS}, 0, 0}
\end{equation}
We wish to show that it is this relation which is preserved by the evolution
equations.  Note that at time $ t = 0 $, we have 
$ z_{l_S, l_{NS}, \tilde{l}_{NS}} = \frac{1}{2} \delta_{l_S + l_{NS}, 0} $,
so that this relation holds at $ t = 0 $.  If we can show that if this
relation holds for all $ z_{l_S, l_{NS}, \tilde{l}_{NS}} $ at some time
$ t $, then it holds for $ d z_{l_S, l_{NS}, \tilde{l}_{NS}}/dt $, it follows
that it holds throughout the evolution.

We note also that $ \kappa_{l_S, l_{NS}, \tilde{l}_{NS}} $ only depends on 
$ l_S + l_{NS} $ in the limit of infinite sequence length.  Therefore, we 
may define $ \kappa_{l_S + l_{NS}} = \kappa_{l_S, l_{NS}, \tilde{l}_{NS}} $, 
with $ \kappa_0 = k $, and $ \kappa_l = 1 $ otherwise.  So, suppose at some 
time $ t $ we have that Eq. (A14) holds for all $ l_S, l_{NS}, \tilde{l}_{NS} 
$.  Then, after switching notation from $ l_S, l_{NS}, \tilde{l}_{NS} $ to 
$ p, q, r $, we have,
\begin{widetext}
\begin{eqnarray}
\frac{d z_{p,q,r}}{dt} 
& = &
\sum_{j = 0}^{p} \sum_{k = 0}^{q - r} \sum_{l = 0}^{r}
\frac{1}{j! k! l!} (\frac{\mu}{4S})^{j+k+l}
(2S-2)^l \kappa_{p+q-j-k-l} {{p+q-j-k-l}\choose{p-j}}{{q-k-l}\choose
{r-l}} \times \nonumber \\
&   &
(2S-2)^{r-l} z_{p+q-j-k-l, 0, 0}
- (\kappa_{p,q,r} + \bar{\kappa}(t)) {{p+q}\choose{p}}{{q}\choose{r}}
(2S-2)^r z_{p+q,0,0} \nonumber \\
& = &
{{p+q}\choose{p}}{{q}\choose{r}}(2S-2)^r \sum_{m = 0}^{p+q}
\frac{1}{m!} (\frac{\mu}{4S})^m \kappa_{p+q-m} z_{p+q-m, 0, 0} 
\times \nonumber \\
&   &
[\frac{1}{{{p+q}\choose{m}}} \sum_{j+k+l = m, (j, k, l) \in [0, p] \times
[0, q-r] \times [0, r]}{{p \choose j}{q-r \choose k}{r \choose l}}]
- (\kappa_{p, q, r} + \bar{\kappa}(t)) {{p+q}\choose{p}}{{q}\choose{r}}
(2S-2)^r z_{p+q,0,0} \nonumber \\
& = &
{{p+q}\choose p}{{q} \choose r} (2S-2)^r \frac{d z_{p+q, 0, 0}}{dt}
\end{eqnarray}
\end{widetext}
The last two lines are derived by noting that the product of the factorials,
$ \frac{1}{j! k! l!}{{p+q-j-k-l}\choose{p-j}}{{q-k-l}\choose{r-l}} $,
is equal to, $ {{p+q}\choose p}{{q}\choose r} \frac{1}{(j+k+l)!} 
\frac{{p \choose j}{q-r \choose k}{r \choose l}}{{{p+q}\choose{j+k+l}}} $, and 
then by noting that $ \sum_{j+k+l = m, (j, k, l) \in [0, p] \times
[0, q-r] \times [0, r]}{{p \choose j}{q-r \choose k}{r \choose l}} =
{{p+q}\choose m} $.  This relation can be derived by expanding 
$ (x+1)^{p+q} $ in two different ways:  First by direct expansion using
the binomial theorem, and second by expanding $ (x+1)^p $, $ (x+1)^{q-r} $,
$ (x+1)^r $ separately, and then taking the product.  Matching powers of
$ x $ yields the relation given above.

Note that that we have shown that $ z_{p, q, r} = {{p+q}\choose p}
{{q}\choose r} (2S-2)^r z_{p+q, 0, 0} $ for all $ p, q, r $ throughout
the evolution.  Then given some $ l $, let us collect all the population
at Hamming distance $ l $ from $ \sigma_0 $ by defining $ w_l =
\sum_{m = 0}^{l}\sum_{r = 0}^{m}{z_{l-m, m, r}} $.  We then have,
$ w_l = z_{l,0,0} \sum_{m = 0}^{l}\sum_{r = 0}^{m} {{l \choose m}{m \choose
r}} (2S-2)^r = (2S)^l z_{l, 0, 0} $.  Therefore, using the expression
for $ dz_{l, 0, 0}/dt $, we immediately obtain the infinite sequence
length equations given by Eq. (17).

\section{Notes on the Implementation of the Stochastic Simulations}

To allow for independent verification of the semiconservative error 
catastrophe, we develop a stochastic framework that directly simulates the 
population dynamics of evolving organisms.  These simulations are stochastic 
in that replication and error events occur with some probability at each time 
step in the simulation.

Although the stochastic system we develop mirrors the system described in the 
semiconservative quasispecies equations, we do not include any {\it a priori} 
information from these equations in our simulations.

The stochastic system consists of a population of $ N $ organisms.
Each organism contains a genome with two strands:  $ \sigma $ and its 
complement $ \bar{\sigma} $.  Each genome sequence is associated with a 
probability of replication $ p_{R, \{\sigma, \bar{\sigma}\}} $ that describes
the probability that an individual with that genome will replicate at each 
time step.  If a genome is chosen to replicate, the strands separate into two 
daughter organisms which proceed to replicate the opposing strand.
Each base on that strand is correctly replicated with a probability of 
$ 1 - \epsilon $.  After complimentary strand synthesis, lesions in the genome 
of the daughter organisms are repaired, and in the absence of 
parent-strand-specific repair mechanisms the ``correct'' base-pairing is 
retained in the daugher with probability $ 1/2 $.

In order to maintain a relatively constant population size in our system, we 
impose the constraint that $ N $ is less than or equal to some $ N_{max} $.
To meet this constraint, after a round of replication proceeds through the 
population, if $ N > N_{max} $, individuals are removed from the population 
until $ N = N_{max} $.  Each individual in the population has equal 
probability ($ 1/N $) of being removed at this step.

Although we employ a single-gene, single-fitness-peak model for the purposes 
of this work, the stochastic framework described here is very general and may 
be employed to explore the dynamics of populations in much more complicated
systems.

\end{appendix}

\bibliography{semi_conserv_ps}

\end{document}